# Leak Proof CMap; a framework for training and evaluation of cell line agnostic L1000 similarity methods

Steven Shave[1,2], Richard Kasprowicz[2], Abdullah M Athar[3], Denise Vlachou[2], Neil O Carragher[1], and Cuong Q Nguyen[4*]

[1]Edinburgh Cancer Research, Cancer Research UK Scotland Centre, Institute of Genetics and Cancer, University of Edinburgh, Crewe Road South, Edinburgh, EH4 2XR, UK. Email: neil.carragher@ed.ac.uk

[2]GSK Medicines Research Centre, Stevenage, SG1 2NY, UK.

[3]Artificial Intelligence and Machine Learning, GSK, The Stanley Building, London, N1C 4AG, UK.

[4]Artificial Intelligence and Machine Learning, GSK, 5th Floor, Suite 1, 259 E. Grand Ave, South San Francisco, California, 94080, USA. Email: cuong.q.nguyen@gsk.com

* Corresponding author, to whom correspondence should be addressed.

## ABSTRACT GRAPHIC / TOC GRAPHIC

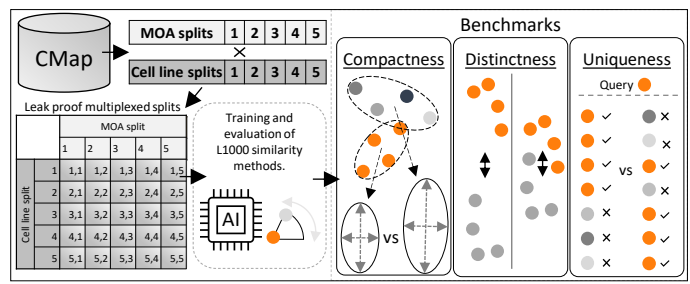

## ABSTRACT

The Connectivity Map (CMap) is a large publicly available database of cellular transcriptomic responses to chemical and genetic perturbations built using a standardized acquisition protocol known as the L1000 technique. Databases such as CMap provide an exciting opportunity to enrich drug discovery efforts, providing a 'known' phenotypic landscape to explore and enabling the development of state of the art techniques for enhanced information extraction and better informed decisions. Whilst multiple methods for measuring phenotypic similarity and interrogating profiles have been developed, the field is severely lacking standardized benchmarks using appropriate data splitting for training and unbiased evaluation of machine learning methods. To address this, we have developed 'Leak Proof CMap' and exemplified its application to a set of common transcriptomic and generic phenotypic similarity methods along with an exemplar triplet loss-based method. Benchmarking in three critical performance areas (compactness, distinctness, and uniqueness) is conducted using carefully crafted data splits ensuring no similar cell lines or treatments with shared or closely matching responses or mechanisms of action are present in training, validation, or test sets. This enables testing of models with unseen samples akin to exploring treatments with novel modes of action in novel patient derived cell lines. With a carefully crafted benchmark and data splitting regime in place, the tooling now exists to create performant phenotypic similarity methods for use in personalized medicine (novel cell lines) and to better augment high throughput phenotypic screening technologies with the L1000 transcriptomic technology.

## ABBREVIATIONS

AI/ML, Artificial Intelligence and Machine Learning; CMap, Connectivity Map; HCI, High Content Imaging; HPC, High Performance Computing/Compute; MOA, Mechanism of Action; NME, New Molecular Entities; PCA, Principal Component Analysis;

### Glossary of terms

Phenotypic similarity method; any metric operating on their phenotypic features (optionally employing an embedding step) used to compare phenotypes, e.g., Euclidean and cosine distance. Split A,B; denotes a combined cell line and MOA Leak Proof CMap split containing a test set with cell line split A cell lines, and MOA split B cell lines (i.e., split 1,2 denotes cell line split 1, MOA split 2).

## INTRODUCTION

Phenotypic screening[1-3] is used with great effect against a backdrop of falling drug approval rates[4-6], pushing efforts towards novel first-in-class therapies and tackling previously dubbed "undruggable" targets[7] in the search for efficacious treatments. Analysis shows that most first-in-class drugs approved between 1999 and 2008 originated from such campaigns[6], hitting targets and possessing mechanisms of action (MOA) undiscoverable with traditional target-based approaches[8]. Indeed, a recent meta-analysis by Sadri suggested that only a striking 9.4 % of approved drugs were discovered using target-based assays[9]. With artificial intelligence and machine learning (AI/ML) impacting every aspect of modern life, it is then unsurprising to see it applied in almost every niche of the drug discovery process[10, 11], and prominently deployed within modern phenotypic screening[12-14]. Key areas being addressed with AI/ML methods are classification of cell phenotypes, target deconvolution and MOA assignment[15-18], which aim to associate responses to potentially novel treatments to phenotypes of targets and pathways. Whilst notionally simple, the polypharmacology of compounds even with a singular clear and strongly affected primary target makes this difficult. One study found that only 0.5 % of individual protein-disease pairings were causal and used this rate to calculate expected clinical failure rates in close agreement with *a priori* estimates[19]. Whilst polypharmacology complicates deconvolution and MOA determination, it is often highlighted as a route to rescue falling drug approval rates[20, 21]. Indeed, numerous existing approved drugs achieve efficacy through their polypharmacological profiles and multiple mechanisms, not exquisite selectivity to a single target[8, 22]. This results in many disease states being treated with a 'magic shotgun' and not a 'magic bullet' approach[23, 24]. However, harnessing such polypharmacology-driven efficacy and even choosing molecules which exploit it must be traded off against the expectation of more difficult MOA assignment, requiring more rigorous treatment of data and the potential to encounter more numerous and novel toxicity mechanisms[25-27] with multiple modalities[28-30]. This presents an opportunity and need to rapidly build new drug discovery tools using AI/ML techniques capable of detecting not just the perturbation of one target or pathway but multiple, revealing drivers of efficacy and flagging potentially undesirable off-target effects. Generalized, hypothesis-free MOA assignment (e.g., without specific markers) typically relies on the existence of a corpus (or ground truth set) of well described and understood treatments spanning desirable (therapeutic) and undesirable (toxic, disease-state promoting, and damaging) MOAs. This may then be searched for close similars using a variety of techniques spanning different transformations, embeddings, and distance/similarity measures to infer MOAs[31].

An assay technology well suited to comparison of newly collected data to a corpus of existing data is L1000[32], a transcriptomic technique using bead-bound probes to capture and measure mRNA for 978 "landmark" genes and 80 control genes present in a lysed cell population. From these landmark genes, a further 9,196 may be confidently inferred. The technique is also accompanied by a valuable asset in that it is backed up by the NIH Library of Integrated Network-Based Cellular Signatures[33] initiative supported Connectivity Map (CMap) database[32, 34], a large public repository of more than 1.3 million L1000 profiles against a range of cell lines treated with chemical and CRISPR perturbations. Despite calls for caution around the reproducibility of CMap profiles from Lim and Pavlidis[35], the database has seen constant use in research as evidenced by the many thousands of citations received by the original publication[34].

## L1000 similarity methods

For simplicity, we use the term phenotypic similarity method to refer to methods operating on L1000 profiles such as AI/ML-derived models, correlation, distance, and similarity measures both with and without transformation and embedding steps. The creation and evaluation of phenotypic similarity methods applied to the CMap database is a well explored topic in literature, with correlation-based phenotypic similarity methods defined in the original CMap publication[34] and later expanded upon with the creation of the commonly used Zhang metric[36] amongst others[37]. Metric learning[38, 39] efforts have highlighted the performance gains possible when operating on large datasets such as CMap, which provides data at a scale suitable for a variety of algorithms and techniques. One approach named perturbation barcodes[40] by Filzen et.al. trained a neural network to minimize the Euclidean distance between MOA matched treatments within a 100-dimensional binary output embedding. Whilst transferability of the model is demonstrated by application to CMap data, it is unknown what compounds or modes of action were present in training, validation, and test sets of the inhouse dataset or how carefully splitting was conducted to keep the same compounds, MOAs, or similars, from leaking from the training to test sets. Additionally, only two cell lines were used to train the model, with no evaluation made as to model transferability across lines.

A similar approach is taken by Donner et.al.[41] in which a deep neural network with 64 densely connected hidden layers is used to embed L1000 profiles using contrastive learning onto a hypersphere, from which embeddings can be compared using cosine similarity. One test set comprises random 20 % splits of perturbagens (small molecule, genetic and both), risking the leak of information from training to test sets by virtue of similar MOAs being present in both. One evaluation mode sees random splitting abandoned and training with all CMap data followed by testing on an external evaluation set containing different perturbagens than those used in training, which again, does not control for MOAs or cell lines shared between training and test sets. Literature documents multiple studies which train neural networks and apply other machine learning techniques to CMap or L1000 data, an overview of which is given by Issa et. al.[42]. Critically, previous approaches to MOA prediction fail to apply rigorous data splitting best practices. This could result in the leak of training data including compounds, MOAs, and cell line specific behavior into validation and test data. Assessing performance in these situations inevitably returns unrealistic and inflated performance scores unachievable with unseen novel treatments and cell lines. In addition, many techniques use internal and non-publicly available datasets making repetition and comparison of approaches difficult[43-46].

In order to realize the full potential of the L1000 technique and CMap in supporting phenotypic screening efforts, metrics must remain performant when matching MOAs from different cell lines, allowing the wealth of information within CMap to be applied to novel and disease relevant cell lines as often seen used in phenotypic screening campaigns[31, 47, 48]. To our knowledge, no fully fledged study has been made in this area. In addition, existing methods fall short in that they are not evaluated using standard best practices, do not make their testing data available, were evaluated on only a few cell lines, and finally, are often evaluated using different benchmarking tasks and performance measures making comparisons difficult.

## The Leak Proof CMap package

Following real-world scenarios where phenotypic similarity methods are deployed, we would like to evaluate their performance and generalizability in three important scenarios: i) novel cell lines, ii) novel MOAs, and iii) novel MOAs in novel cell lines. To achieve this, Leak Proof CMap defines a rigorous data splitting scheme designed to be leak proof, and ensure that no similarly responding cell lines or MOAs leak from training, validation or test sets[49]. This is achieved using a protocol summarized in Figure 1, generating five intermediate MOA splits and five intermediate cell line splits with the goal of pushing similar lines or MOAs into the same split and producing maximum diversity across splits.

## A) Splitting cell lines and MOAs

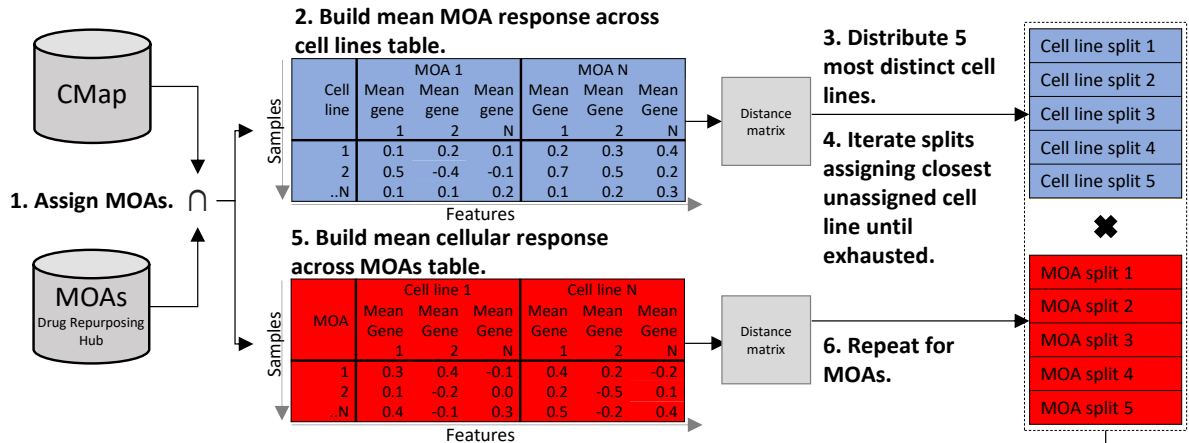

## B) Combining cell line and MOA splits into Leak Proof CMap splits

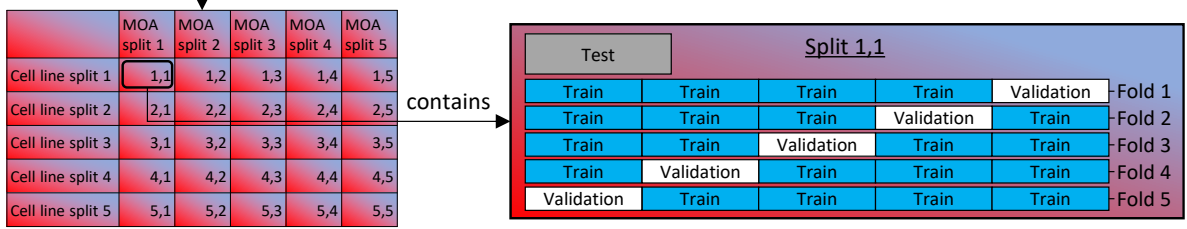

*Figure 1 - Data splitting regime. A) Splitting cell lines and MOAs into 5 cell line and 5 MOA splits using the CMap database and compound MOA information from the Drug Repurposing Hub. Two tables are generated (with missing values imputed) representing; i) mean MOA response across cell lines, with rows representing cell lines and feature columns containing gene Z-scores for each MOA, and ii) mean cellular response to MOA treatments, with rows representing MOAs, and feature columns containing gene Z-scores for each cell line. The 5 most diverse cell lines from table i are placed into a unique MOA split, which are then filled in a round-robin manner assigning cell lines to splits which minimally increase all-to-all inter-split distances. The same protocol is applied to assign MOA splits using table ii. B) Combining cell line and MOA splits into Leak Proof CMAP split objects creates 25 combined splits. These combined splits hold out their cell line split and MOA split data in a test set, and are assigned names denoting this, such that split 1,2 contains held out test data from cell line split 1, and MOA split 2. Every Leak Proof CMap split contains held out test data along with the remaining data assigned to 5-folds of training and validation data.*

We enumerate combinations of these splits to arrive at a final set of 25 Leak Proof CMap splits (five cell line splits for each of five MOA splits), each representing a held out test set of L1000 profiles from their original cell line and MOA splits. In addition to the held out test sets, 5-fold training and validation sets are also captured (comprising all data not within the test set) for model training, selection and validation. This ensures performance evaluation can be conducted against trained models with confidence that no information from similar cell lines or MOAs is leaked from training or validation into test sets; closely simulating practical target deconvolution and MOA determination for treatments on novel and potentially patient-derived cell lines.

With test data defined, Leak Proof CMap evaluates phenotypic similarity methods using three carefully chosen benchmark tasks as shown in Figure 2, evaluating compactness, distinctness, and uniqueness of replicate groups (see experimental section).

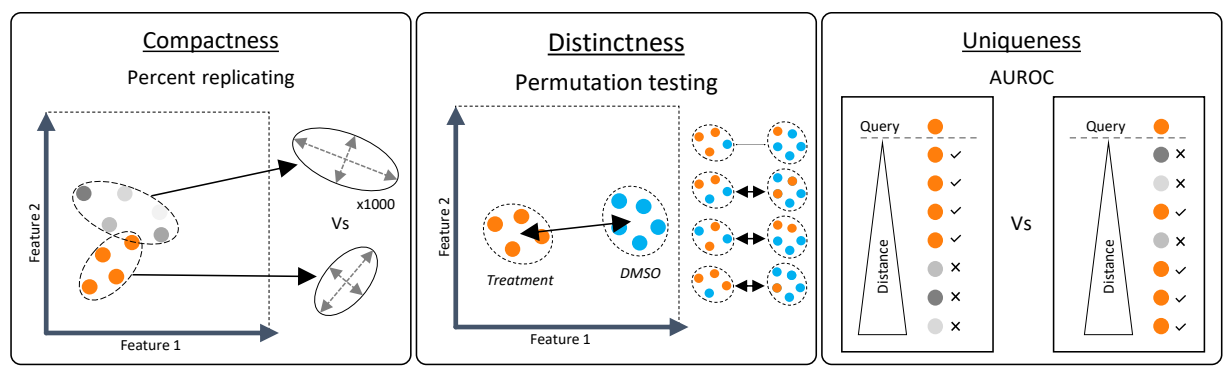

*Figure 2 – The Leak Proof CMap package evaluates phenotypic similarity methods in three benchmark tasks, each with 4 different replicate group criteria covering evaluation within and across cell lines, and matching molecules and MOAs. The three benchmark tasks are; i) compactness, as evaluated using the percent replicating metric which compares the spread of replicate treatments to a treatment background, ii) distinctness; whereby permutation testing is used to determine treatment separation from DMSO controls, and iii) uniqueness, using the AUROC metric to evaluate replicate recall rankings.*

These three benchmark tasks were deemed to capture the most important characteristics of a phenotypic similarity method applied to HTS campaigns running at scale. See supporting information section "Compactness vs Distinctness vs Uniqueness" and Figures S4 to S10 for an intuitive interpretation of these benchmarked characteristics. We use performance in these tasks to define a standard with which to compare all phenotypic similarity methods.

In addition to providing data splits, Leak Proof CMap facilitates training of phenotypic similarity methods using prescribed splits and subsequent evaluation of new and existing phenotypic similarity methods using the three benchmark tasks. Leak Proof CMap was used to train and evaluate the following phenotypic similarity methods: i) "TripletLoss"; included to exemplify the ease in which new AI/ML-derived methods can be trained and evaluated using Leak Proof CMap. Created using a neural network and trained using triplet loss-based methods, the model embeds L1000 profiles into a 128-dimensional space, which may then be compared to other embeddings using cosine distance (see experimental section). The model is designed to match MOAs across any cell line, being trained with triplets composed across cell lines and perturbation treatment concentrations. ii) "Rank" a common and simple similarity method used for comparison of L1000 profiles, whereby the ranked gene transcription levels are compared through application of the Spearman's rank correlation coefficient. iii) "Zhang"[36], a commonly used similarity method for comparison of L1000 profiles, created with the goal of being more robust and performant than methods described in original CMap literature[34]. iv) "Cosine" a common distance metric in AI/ML literature, referring to measurement of cosine distance between two L1000 profiles after application of standard scaler and principal component analysis capturing 99.5 % variance and fitted to an appropriate training set (see experimental section). v) "Euclidean" the Euclidean distance between L1000 level 4 profiles. vii) "EuclideanPCA" the Euclidean distance between two L1000 profiles after application of standard scaler and principal component analysis capturing 99.5 % variance and fitted to an appropriate training set (see experimental section).

Inclusion of the TripletLoss-derived similarity method is intended to show the power and performance gains possible with application of AI/ML techniques. Moreover, it highlights the modular structure of Leak Proof CMap that confers ease of application to a variety of newly developed phenotypic similarity methods, which may be trained and evaluated with appropriate best practices and rigor.

## EXPERIMENTAL SECTION

All source code is available at https://github.com/GSK-AI/leak-proof-cmap and contains the Leak Proof CMap package (with the python package adopting the short name 'leakproofcmap'), along with the notebooks used during the course of this work and in figure production for manuscript preparation. Computation was performed by submission to Linux HPC resources or interactive sessions for the running of Jupyter notebooks. Single Nvidia 3090 GPUs were requested for jobs where GPU acceleration was required. All code ran in Python 3.10.4, using the packages defined as requirements for Leak Proof CMap.

### Dataset preparation – download and down sampling

CMap data was obtained using the Leak Proof CMap package which in turn uses functionality within the Phenonaut[14] (v1.3.4) package to download and process CMap level 4 data capturing Z-scored gene expression counts. Calling Leak Proof CMap functions which rely on this data will trigger download and processing if absent. Additionally, the clue.io Drug Repurposing Hub[50] Drug Information dataset (release version 2/24/2020) is automatically obtained by Leak Proof CMap, which contains information on compounds including identifiers and MOAs. All compounds within CMap not assigned a MOA through matching of their 'pert_iname' fields present in each dataset were removed. To ensure the dataset is not skewed by repeats of special interest and control compounds, the CMap pool is down sampled ensuring a maximum of 200 profiles for each unique treatment are kept. The negative control DMSO is a special case which we would like to include in all training sets for fair and equal comparison so is allowed to remain in the compound pool with no assigned MOA but is limited to 200 samples per cell line present; thus, each cell line in the dataset has up to 200 negative control replicates for comparison. The lowest number of profiles for a unique compound was found to be bruceantin (71 profiles). The IDs of included profiles may be found in the Leak Proof CMap source repository and is named 'downsampled_cmap_identifiers_v1.csv'. This data is then used to instantiate a Phenonaut object wrapping the data and capturing features (978 genes) for ease of handling. These 177,369 profiles represent 1,309 unique compounds at multiple doses across 30 cell lines with 433 unique MOAs as defined by the Drug Repurposing Hub. A breakdown of MOA samples per cell line is available in the supporting information spreadsheet file accompanying this manuscript.

## Dataset preparation – defining data splits

The Jupyter notebook 'l1000bench_00_define_splits.ipynb' in the Leak Proof CMap source repository captures all steps used to generate appropriate train, validation, and test splits. Code within the notebook splits the CMap database so that model performance may be evaluated using data absent from training whilst ensuring that highly similar data is not present across training/validation and test sets. Data splitting is performed with respect to both cell lines and MOAs aiming to define five splits for each, keeping similars within splits and diversity between splits. Targeting five splits strikes a balance between the amount of data within each split and the number of test set evaluations required after pairing every cell line split with every MOA split. Figure 1 shows an overview of the initial split creation process and final split combinations in order to produce 25 Leak Proof CMap splits. For initial cell line splits, mean profiles are generated for each MOA. However, profiles for all MOAs are not present in all cell lines, with only six MOAs found in all 30 cell lines, ('EGFR inhibitor', 'mTOR inhibitor', 'AKT inhibitor', 'MEK inhibitor', 'PI3K inhibitor', and 'CDK inhibitor'). We therefore used the KNNImputer from Phenonaut (v1.3.4) which wraps Scikit-Learn's[51] KNNImputer to impute missing values, reviving the MOA coverage to the full complement of 433 MOAs across 30 cell lines. Next, all-to-all cosine distances for cell line averaged Z-scores of gene counts in response to all MOAs are calculated. This distance matrix is then used to identify the top five most diverse cell lines by virtue of them having the highest average distance to all other cell lines (in MOA response space). Once these five diverse cell lines are added to splits, remaining cell lines are assigned splits by iterating splits and finding the closest cell line not within a split to any member of the split. This continues until all cell lines are assigned to splits, generating diverse splits full of similar cell lines. The chosen splitting solution maximizes similarity and diversity whilst maintaining suitability for assigning a large number of items to splits and also suitability to apply the same technique to both cell line and MOA split creation; important factors, which are unattainable with alternative solutions e.g. exhaustive enumeration. The same procedure was conducted to define MOA splits, first imputing absent values where no compound with the required MOA was evaluated in each cell line, and then applying the same split assignment algorithm to the pool of 433 MOAs. Cell line and MOA splits can be found in the supporting information accompanying this manuscript.

With five cell line splits and 5 MOA splits, combined Leak Proof CMap splits are created by pairing each cell line split with each MOA split, defining 25 splits which dictate the cell lines and MOAs to be held out of training and validation sets and used for the testing and performance evaluation of models only. Throughout this manuscript we refer to splits by their two categorical split numbers with cell line split first, followed by a comma and the MOA split number. Therefore, "split 1,2" denotes cell line split one, MOA split two. The Leak Proof CMap package provides functionality for the creation of CMapSplit objects which supply requested training, validation, and test data for a given split. Data may also be requested in a manner which returns folds of training and validation data for use in cross fold validation. Furthermore, the option exists to return data which has undergone application of mean and variance scaling and also transformed into principal component space, by fitting both scaling and PCA transformers to the training data of split 1,1 in the first of the five folds.

## Training and model selection

The Leak Proof CMap package implements a flexible metric learning neural network using triplet loss and built with the PyTorch Lightning[52] framework (v 2022.10.25). The TripletMarginWithDistanceLoss function from PyTorch, utilizing cosine distance was used as the loss function for model training which takes the form : $l_i = \max[\, d(a_i, p_i) - d(a_i, n_i) + m, 0]$, where $l_i$ is the loss for triplet $i$, $d$ is the cosine distance function, $a$ is the anchor profile, $p$ is the positive (similar) profile, $n$ is the negative (different) profile, and $m$ is the margin term, added to the difference of distances. The margin term allows ignoring of triplets performing well enough through exclusion of their likely negative distance differences and allows the most incorrect triplets to influence training. A margin value of 0.2 was chosen empirically and represents 10 % of the range of the cosine distance metric. The number of hidden layers within the network is flexible, along with the number of nodes within each layer. Between each layer of the network, there is a batch normalization operation, and a dropout layer that the ReLU activation function is applied to. An exception to this is the last layer, which is an embedding layer, where newly embedded data representative of a transformed L1000-derived phenotype (anchor) will be represented and compared against matching profiles (positive) and non-matching (negative) profiles during training. It is this comparison of an anchor to positive and negative profiles that allows for learning of powerful embeddings invariant to batch effects present in the anchor and positive samples. We exploited this phenomenon to train a neural network invariant to cell line and compound concentration by presenting positives (matching compound treatment) from different cell lines as similar to an anchor from another, and at different concentrations. This triplet forming scheme was achieved using a dataloader which forms triplets across cell lines and does not consider treatment concentration information, matching only unique compound treatments for the anchor and positive samples. Hyperparameter scanning was performed using the Optuna[53] (v 3.3.0) framework, optimizing the number of layers in the neural network, number of nodes within each layer (different for each layer), output embedding size restricted to powers of 2, initial learning rate, batch size, and dropout rate. Early stopping was applied within each model, monitoring validation loss with a patience of three epochs, and run for a maximum of 300 epochs. All code and hyperparameter optimization was run multiple times to ensure reproducibility with initial random states supplied to Numpy[54] (v 1.24.4), PyTorch[55] and Optuna random number generators used in model initialization, training and evaluation. All hyperparameter scanning was conducted by training on the training set of split 1,1, with validation accuracy calculated on the validation set of fold 1, which reports the accuracy rate achieved in selection of matching compound profiles against a randomly chosen non-replicate profile compound treatment. Each repeat of the hyperparameter scan used 48 hours of HPC time on one GPU node to ensure reproducibility. Replicate runs agreed exactly on model performance and hyperparameters explored, although the first run completed 533 trials (see supporting information accompanying this manuscript for Optuna trial results, hyperparameters and model metrics), and the second completed 457 trials, with the difference in the number of trials complete down to the HPC environment including network, interconnects, and storage load. Trial 155 produced the model with the highest validation accuracy score of 0.8728 and also the lowest validation loss of 0.0619. See Table S1 for hyperparameters and spreadsheet worksheet 'OptunaResults' for full hyperparameter scan results, both of which may be found in supporting information accompanying this manuscript. From this point on, all trained models used hyperparameters from trial 155. 5-fold cross validation was performed using the folds defined by the CMapSplit object for split 1,1 and repeated with five different initial random states to assess model stability. Figure S1 in supporting information shows the 5-fold cross validation performance of models trained with different initial random seeds, all of which are stable and performant. A one-way ANOVA test showed no significant difference between these cross fold validation scores (P value = .69). Validation sets were then merged into training sets for each of the 25 splits (defined in Dataset Preparation above and in Figure 1) and a new model trained using this data for each split over 16 epochs (the median number of epochs reached in cross fold validation). Model checkpoints were saved and used to evaluate held out test set accuracies. Figure S2 shows these test set accuracies for each split trained across five different initial random seeds. A decision to use an initial random seed of seven for subsequent phenotypic similarity method evaluation was taken before generating this data so as not to bias performance. Models trained on splits with an initial random seed of seven achieved an average test split accuracy of 0.8290 with a standard deviation of 0.0181 indicating close performance across splits.

During model training, triplets may be placed into three categories; hard, semi-hard and easy. Training triplet loss-based models on semi-hard triplets defined by the negative being further away from the anchor than the positive but within the margin (producing a positive loss), is known to most efficiently train performant models[56]. The triplet types comprising batches encountered during training of the chosen model on split 1,1 were monitored. It was observed that the fraction of semi-hard triplets within a batch converged to around 0.2 by completion of model training with notable shifts of hard to semi-hard, to easy triplets. Supporting information Figure S3 shows triplet fractions encountered in each training batch. Code used to perform this analysis is available in the Jupyter notebook 'lpcmap_02_investigate_model_training.ipynb' within the Leak Proof CMap source code repository. Whilst explicit online batch mining is often used to compose semi-hard triplets for training purposes, it is likely that composing triplets across cell lines and treatment concentrations naturally forms semi-hard triplets and is an effective strategy in data preparation for contrastive learning.

### Replicate group criteria

Use of different replicate group criteria enables more thorough evaluation of phenotypic similarity methods. Evaluation of groups comprised of identical treatments of the same concentration within the same cell lines differs greatly from a grouping in which replicate groups are formed by any treatments sharing common MOAs, at any concentration, and in any cell line. Phenotypic similarity methods performing well in this second replicate grouping criteria would be of great use in target deconvolution and MOA assignment for new treatments in novel patient derived cell lines in contrast to the first. We have therefore created four different replicate grouping criteria which when suffixing the benchmark tasks compactness, distinctness, and uniqueness, indicate the criteria used to form replicate groups. We illustrate their use with the compactness benchmark; i) Compactness of treatments within cell lines, referred to as simply 'Compactness'. Highly similar to the original definition of percent replicating (see below), we define replicates to comprise the same compound at the same concentration in the same cell line and build background non-replicate groups (required for compactness) comprised of unique non-matching compounds at the same concentration and in the same cell line as the replicate treatment. ii) Compactness of compounds across lines, referred to as 'Compactness across lines' for brevity. Replicate treatments match compound at any concentration and are taken from all cell lines, non-replicate groups are formed by taking unique non-replicate matching compounds from a single cell line which contributed to the replicate group. iii) Compactness of MOAs within cell lines, referred to as 'Compactness MOAs' for brevity. Similar to 'Compactness', but matching treatment MOAs instead of compounds or concentrations and composing non-replicates from unique MOAs within the matching replicate cell line. No concentration information is considered in this mode. iv) Compactness of MOAs across lines, referred to as 'Compactness across lines MOAs' for brevity. Replicates have matching MOAs, but are drawn from different cell lines, non-replicate groups comprise non-replicate matching unique MOAs drawn from a cell line which contributed to formation of the replicate group.

## The compactness benchmark task

The compactness benchmark (Figure 2, left) task evaluates how well replicate treatment groups group together with respect to a background of non-replicate treatments. The ideal phenotypic similarity method would overlap replicates, controlling for batch effects, different cell lines, and concentrations, whilst a poorly performing one would scatter replicates widely and overlap with distinct non-replicate treatments. The Percent Replicating[57] metric was chosen as an appropriate measure for compactness evaluation, providing a simple percentage readout of the number of treatment replicate groups more compact than the 95th percentile of a background comprising distinct non-replicate treatments. Other approaches to measuring compactness such as Silhouette score[58], and simpler distance-based measures were considered but rejected due to not meeting important constraints on operating on broad types of data with diverse embeddings, and with any phenotypic similarity method. Percent replicating met these criteria after a small modification (described below) allowing it to work with any distance or similarity-based phenotypic metric and not just the literature defined Spearman's rank correlation coefficient. The percent replicating metric described by Way et.al.[57] and implemented in Leak Proof CMap was run using the four different replicate group criteria defined above, using all test splits for all similarity methods, loading the appropriate model checkpoint for the triplet-loss derived metric.

In all replicate group criteria, percent replicating was run using the default recommendation of comparing median replicate distances with a background of 1,000 median non-replicate distances. The Leak Proof CMap percent replicating implementation returns information on: (i) the percentage of molecules deemed replicating by comparison of distance-based phenotypic metrics to the $5^{th}$ percentile of a background distribution comprising median non-replicate all-to-all distances, and similarity/rank-based phenotypic metrics to the $95^{th}$ percentile; (ii) extended statistics, from which is it possible to test treatments at multiple percentile cut-off levels, allowing creation of plots and tables showing percent replicating verses different percentile cut-offs. For ease of comparison, the results of distance metrics are transformed into a form comparable with similarity metrics which enables comparing all metrics to common cut-offs such as the $95^{th}$ percentile. This is achieved by simply negating distances for distance metrics. We report percent replicating scores according to two analysis modes. The first calculated by averaging individual percent replicating scores achieved for each split along with standard deviations. For the second mode, we also calculate merged split scores which are the result of merging treatment group replicating status from all splits and reporting the overall percent deemed replicating in Table S2 within supporting information accompanying this manuscript. The significance of performance differences between percent replicating values are calculated using the Wilcoxon signed rank test (see experimental section). Merged split performance significances were assessed using McNemar's test as implemented in the statsmodels[59] (v 0.14.0) Python package, operating on the binary assignment of replicating or non-replicating status for matched samples across phenotypic similarity methods (see supporting information accompanying this manuscript). In addition to the two analysis modes for the four replicate group criteria detailed above, we also apply a modifier for included compound treatments, in that evaluation is performed on a subset of compounds from the JUMP consortium[60], which include 90 well-studied compounds covering 47 MOAs with high confidence. However, only 25 of these compounds are present in CMap. When benchmarking with these 25 compounds, we refer to them as the 'JUMP MOA compound subset'.

## The distinctness benchmark task

Distinctness from DMSO (or distinctness from vehicle) is an important measure applied to every sample in a HTS assays in an effort known as hit calling (Figure 2, center). Historically, hit calling with univariate readouts used simple thresholds like the three-sigma level away from the mean of negative controls. Nonlinear multiparametric readouts make application of such thresholds difficult, necessitating application of measures within the space which are invariant to axis scale. Such measurements include Mahalanobis distance as applied to multiparametric readouts by Caie et.al.[61] and Hughes et.al.[62] and variants like the Multidimensional Perturbation Value[63] operates on the principal components of merged treatment and DMSO samples. Permutation testing[64, 65] was chosen in the present study to evaluate phenotypic similarity methods in their assignment of treatments being distinct from DMSO. This choice was based upon there being no requirement for normally distributed data in contrast to using Mahalanobis distances and the Multidimensional Perturbation Value. It is a non-parametric test which can operate on any phenotypic similarity method in any embedding space. Permutation testing was carried out using the four replicate group criteria defined above with DMSO being used in place of the non-replicate group. Functionality of Leak Proof CMap was used to perform the test which starts with the null hypothesis that both the treatment and DMSO come from the same distribution. First, the distance from the average treatment in a group to the average DMSO treatment is calculated. Then, with the goal of sampling all possible label permutations of "treatment" vs "control", the binomial coefficient equation is used to determine if enumeration of all groups would exceed 10,000 trials and if so, random sampling of label permutations is used to collect the number of times that average treatment to average DMSO distances were the same or increased under permuted labels verses the original labels. If the total number of permutations is under 10,000 then all are enumerated and evaluated. The correct initial permutation is always included in both sampling and enumeration permutation modes to ensure no P values equal to zero are returned when calculated by dividing the number of equal or greater than distances found between permutation groups by the number of explored permutations. For ease of display of P values for treatment distinctness in figures, we transform P values by taking their negative $\log_{10}$, meaning the output of runs with 10,000 permutations are transformed onto a scale between zero and four, with a threshold of 1.30103 representing the 0.05 alpha cut-off, above which (due to the negation step), the null hypothesis is rejected, and the treatment group deemed distinct from DMSO. Similarly to the compactness task, we report average percent distinct scores denoting the number of compounds which were deemed distinct from DMSO, calculated for all splits and also report standard deviations of these split averages. Distinct from DMSO status is assigned to treatment groups by applying an alpha cut-off of .05. In addition, and similarly to the compactness task, we also calculated merged split results, whereby treatment group distinctness status from all splits was merged into one dataset and then the overall average percent distinct for each technique calculated (see table S2 in supporting information). As with the compactness task, significance is assessed for split averaged performance using the Wilcoxon signed rank test, and for merged splits using McNemar's test (see supporting information methods). Analysis used all CMap filtered compounds as well as the JUMP MOA compound subset.

### The uniqueness benchmark task

The performance of phenotypic similarity methods in retrieval of matching profiles is evaluated in the uniqueness task, calculating Area Under the Receiver Operating Characteristic (AUROC) curve values which quantify how highly ranked replicate treatments occur by distance to a matched replicate treatment verses non-replicate treatments (see supporting information for intuition around the relation of AUROC scores and uniqueness, and Figure 2, right). Receiver Operating Characteristic (ROC) curves[66] provide a measure of how highly ranked true positives are as a function of all other ranked elements in the prediction set when applying a range of cut-offs. These curves may then be further reduced to a single number by calculation of Area Under the ROC curve (AUROC) using a variety of techniques. Replicate groups were defined using the four replicate grouping criteria defined above. Careful creation of null or background distributions was not required as these were formed from all other available treatments (and replicates) within test splits. As treatment groups comprise multiple replicates, each replicate was used as a query profile and then AUROC scores generated from the ranked recall of other replicates within the group, after which, these AUROC scores were averaged to obtain an average AUROC score for a replicate group. Within splits, these AUROC scores for different replicate groups were averaged to assign an AUROC score to the split, after which this score was averaged for all splits and reported as a split averaged AUROC score along with standard deviations (see Table 2). We also report a merged split AUROC/uniqueness score, which is the result of calculating average treatment group AUROC scores and merging this information from all splits before performing an average and standard deviation calculation for all merged split AUROC scores. Statistical significance testing was conducted on both split averaged and merged split AUROC scores using the Wilcoxon signed rank test. Analysis of metrics used all filtered CMap treatments, along with the JUMP MOA compound subset as defined in the compactness benchmark task experimental section.

## Statistical testing

### Kruskal-Wallis H-test

The Kruskal-Wallis H-test as implemented in the SciPy[67] (v 1.11.1) Python package was used to assess the presence of significant performance differences between metrics. With a null hypothesis that the median of groups are equal, it is rejected if the calculated P value is greater than the chosen alpha level of .05. Code used to apply this test to split averaged compactness, distinctness, split averaged uniqueness, and merged split uniqueness scores is available in the Leak Proof CMap source repository within Jupyter notebooks[68] named 'lpcmap_03_pctrep_eval.ipynb', 'lpcmap_04_distinct_from_dmso_eval.ipynb', and 'lpcmap_05_auroc_eval.ipynb'.

### Wilcoxon test

The Pairwise Wilcoxon test, also known as the Pairwise Wilcoxon Signed Rank test, was applied both with and without the Benjamini-Hochberg[69] multiple testing correction as implemented in the SciPy[67] (v 1.11.1) Python package, and was used to determine if certain metrics were performing significantly better than others. Code used to apply this test to split averaged compactness, split averaged distinctness, split averaged uniqueness, and merged split uniqueness scores is available in the Leak Proof CMap source repository within Jupyter notebooks[68] named 'lpcmap_03_pctrep_eval.ipynb', 'lpcmap_04_distinct_from_dmso_eval.ipynb', and 'lpcmap_05_auroc_eval.ipynb'. Analysis in this manuscript was constrained to considering only the singular application of tests using the top ranked, and second top ranked metrics so that no false discovery rate correction is required, however, full all-to-all comparisons with the multiple testing correction applied are available in supporting information accompanying this manuscript (See the supporting information spreadsheet sheets named 'Compactnes_P_values', 'Distinctness_P_values', and 'Uniqueness_P_values').

## RESULTS

Leak Proof CMap was used to benchmark phenotypic similarity methods, first evaluating the compactness of replicate groups across a range of replicate group criteria using the percent replicating metric. Next, hit calling was evaluated using the distinctness (from DMSO) task whereby permutation testing of replicate groups is used to determine the ability of phenotypic similarity methods to distinguish treatments from DMSO controls, a common task in the analysis of assay results. Finally, Leak Proof CMap evaluates replicate groups using the uniqueness task, whereby the AUROC metric is used to assess recall of true positives as a function of all other samples present within a test split. To obtain a view of how well each metric, or model trained on a specific split performs, we calculated split averaged compactness, distinctness, and uniqueness scores from the 25 test splits.

Table 1 illustrates the number of profiles, unique MOAs and unique compounds within each split, evidencing an even distribution across all elements.

*Table 1 – Unique profile, MOA, and compound counts for training + validation data, and test data across all 25 splits.*

| Cell line split, MOA split | Training + validation data | | | Test data | | |
|---|---|---|---|---|---|---|
| | Profile count | Unique MOA count | Unique compound count | Profile count | Unique MOA count | Unique compound count |
| 1,1 | 86,176 | 346 | 1,071 | 4,976 | 87 | 238 |
| 1,2 | 89,838 | 346 | 1,059 | 4,744 | 87 | 250 |
| 1,3 | 82,495 | 347 | 1,051 | 5,021 | 86 | 258 |
| 1,4 | 78,672 | 347 | 954 | 6,782 | 86 | 355 |
| 1,5 | 89,438 | 347 | 1,102 | 4,011 | 86 | 207 |
| 2,1 | 97,011 | 290 | 892 | 5,356 | 87 | 238 |
| 2,2 | 101,906 | 288 | 928 | 5,048 | 87 | 250 |
| 2,3 | 91,438 | 288 | 832 | 5,234 | 86 | 258 |
| 2,4 | 88,969 | 293 | 820 | 7,209 | 86 | 355 |
| 2,5 | 100,146 | 285 | 907 | 4,225 | 86 | 207 |
| 3,1 | 86,993 | 346 | 1,071 | 4,914 | 87 | 238 |
| 3,2 | 91,016 | 346 | 1,059 | 4,653 | 87 | 250 |
| 3,3 | 83,548 | 347 | 1,051 | 5,094 | 86 | 258 |
| 3,4 | 79,750 | 347 | 954 | 6,820 | 86 | 355 |
| 3,5 | 90,647 | 347 | 1,102 | 3,957 | 86 | 207 |
| 4,1 | 84,144 | 290 | 892 | 8,277 | 87 | 238 |
| 4,2 | 89,003 | 288 | 928 | 8,614 | 87 | 250 |
| 4,3 | 79,893 | 288 | 832 | 8,922 | 86 | 258 |
| 4,4 | 77,579 | 293 | 820 | 12,346 | 86 | 355 |
| 4,5 | 87,499 | 285 | 907 | 7,155 | 86 | 207 |
| 5,1 | 81,718 | 290 | 892 | 9,187 | 87 | 238 |
| 5,2 | 85,580 | 288 | 928 | 9,149 | 87 | 250 |
| 5,3 | 77,015 | 288 | 832 | 9,654 | 86 | 258 |
| 5,4 | 74,945 | 293 | 820 | 13,145 | 86 | 355 |
| 5,5 | 84,209 | 285 | 907 | 7,683 | 86 | 207 |

In addition, and to provide an analysis free from data splitting artefact influences, we calculated merged split scores (merging compactness and distinctness categorical labels and AUROC scores from all splits). Merged split scores are more akin to the regular performance testing of metrics not requiring training (i.e. Rank, Zhang, and Cosine), and the Euclidean metrics, not requiring careful train-validation-test splitting to ensure training data does not leak into test data or that the test data influences model selection. Merged split scores can be found in Table S2 in supporting information.

Table 1 displays split averaged results from all three benchmark tasks for all replicate grouping criteria, denoted by benchmark task name, benchmark task name "across lines", benchmark task name "moas", and benchmark task name "across lines MOAs", for all CMap filtered compounds, and also evaluated using only the JUMP MOA compound subset.

## Compactness

Evaluation of split averaged percent replicating performance for all filtered CMap compounds using the Kruskal-Wallis H-test identified significant performance differences between metrics for 'compactness' (P value < .001), 'compactness across lines' (P value = .026), and 'compactness MOAs' (P value < .001), but not for 'compactness across lines MOAs' which showed large standard deviations between split performance scores (P value = .26). The same test was applied to the JUMP

MOA compound subset which assigned significant performance differences present in the 'compactness' and 'compactness MOAs' replicate group criteria (P values < .001) but failed when applied to 'compactness across lines' and 'compactness across lines MOAs', due to metrics achieving percent replicating scores of zero. For the All CMap filtered compound set, testing the top ranked metric against the second ranked metric reveals that TripletLoss significantly outperforms Rank in 'compactness' and 'compactness across lines' and 'compactness across lines MOAs' (P values: < .001, .003, and .045 respectively). For the JUMP MOA compound subset, TripletLoss significantly outperforms Cosine in 'compactness' and in 'compactness MOAs' (both P values < .001).

### Distinctness

Performance differences between metrics in the distinctness benchmark task are significant for all replicate groups and compound sets, with all P values < .001 with the exception of 'distinct across lines MOAs' for all filtered CMap compounds (P value = .008), and 'distinct across lines' for the JUMP MOA compound subset (P value = .007). However, the TripletLoss model is only ranked top, and significantly so (P value < .001) once for percent distinct within the All Filtered CMap compound set. Of note is the high performance of the EuclideanPCA metric across modes and compound sets for the percent distinct benchmark.

### Uniqueness

The uniqueness task displays significant differences in metric group performance for all replicate group criteria and compound sets (all P values < .001), with the TripletLoss metric significantly outperforming the second ranked metric in all tests. The full performance table for all benchmarks is shown in Table 2 with significance indicators highlighting when the top ranked metric outperformed the second ranked metric at the .05, .01, and .001 alpha levels.

*Table 2 – Means and standard deviations of split averaged benchmark task scores for compactness (percent replicating), distinctness (percent distinct from DMSO), and uniqueness (AUROC). A \* denotes that the top ranked phenotypic similarity method (bold) significantly outperforms the second ranked method (italics) according to the Wilcoxon signed rank test with a P value of < .05, \*\* denotes a P value of < .01, and \*\*\* denotes a P value < .001. The same P value notation is used on benchmark task names to denote significant differences in method performance were detected using the Kruskal-Wallis H-test.*

All filtered CMap compounds

| Phenotypic similarity method | Compactness (%) *** | Compactness across lines (%) * | Compactness MOAs (%) *** | Compactness across lines MOAs (%) |
|---|---|---|---|---|
| TripletLoss | ***55.217 ± 7.859** | **26.771 ± 26.306 | 66.804 ± 7.935 | *21.423 ± 27.254 |
| Rank | *37.930 ± 5.628* | *16.322 ± 11.195* | 77.209 ± 5.766 | *12.399 ± 10.107* |
| Zhang | 37.319 ± 5.698 | 15.039 ± 10.816 | 76.549 ± 6.280 | 10.974 ± 11.651 |
| Cosine | 35.913 ± 3.274 | 11.679 ± 9.100 | **84.472 ± 4.999** | 9.663 ± 9.802 |
| Euclidean | 12.571 ± 1.111 | 13.175 ± 9.937 | 27.912 ± 4.539 | 9.972 ± 8.405 |
| Euclidean PCA | 11.441 ± 1.235 | 10.214 ± 8.516 | 27.566 ± 4.020 | 8.110 ± 11.004 |

| Phenotypic similarity method | Distinctness (%) *** | Distinctness across lines (%) *** | Distinctness MOAs (%) *** | Distinctness across lines MOAs (%) ** |
|---|---|---|---|---|
| TripletLoss | ***45.500 ± 12.225** | 94.026 ± 9.157 | 74.656 ± 9.652 | 92.861 ± 9.239 |
| Rank | 18.039 ± 10.263 | 91.206 ± 12.352 | 80.142 ± 8.515 | 92.916 ± 7.604 |
| Zhang | 17.286 ± 9.676 | 90.564 ± 12.899 | 79.540 ± 8.925 | 92.405 ± 7.479 |
| Cosine | 15.315 ± 7.370 | *97.579 ± 2.666* | *82.525 ± 4.103* | **97.223 ± 2.596** |
| Euclidean | *20.481 ± 7.093* | 95.207 ± 1.772 | 81.848 ± 4.986 | 94.074 ± 3.490 |
| Euclidean PCA | 15.198 ± 4.890 | **97.958 ± 1.481** | ***88.924 ± 4.437** | *97.036 ± 1.946* |

| Phenotypic similarity method | Uniqueness (AUROC) *** | Uniqueness across lines (AUROC) *** | Uniqueness MOAs (AUROC) *** | Uniqueness across lines MOAs (AUROC) *** |
|---|---|---|---|---|
| TripletLoss | ***0.916 ± 0.011** | ***0.840 ± 0.020** | ***0.775 ± 0.016** | ***0.732 ± 0.014** |
| Rank | 0.823 ± 0.013 | 0.681 ± 0.029 | 0.675 ± 0.008 | 0.622 ± 0.019 |
| Zhang | *0.852 ± 0.020* | *0.697 ± 0.041* | 0.680 ± 0.018 | *0.633 ± 0.028* |
| Cosine | 0.837 ± 0.018 | 0.687 ± 0.043 | *0.692 ± 0.013* | 0.627 ± 0.028 |
| Euclidean | 0.757 ± 0.013 | 0.579 ± 0.015 | 0.610 ± 0.014 | 0.551 ± 0.013 |
| Euclidean PCA | 0.764 ± 0.019 | 0.580 ± 0.024 | 0.595 ± 0.013 | 0.553 ± 0.017 |

JUMP MOA compound subset

| Phenotypic similarity method | Compactness (%) *** | Compactness across lines (%) | Compactness MOAs (%) *** | Compactness across lines MOAs (%) |
|---|---|---|---|---|
| TripletLoss | ***36.376 ± 9.529** | 0.000 ± NA | ***96.616 ± 4.869** | 19.444 ± 30.581 |
| Rank | 16.799 ± 9.186 | 0.000 ± NA | 48.430 ± 21.751 | 0.000 ± NA |
| Zhang | 16.671 ± 9.583 | 0.000 ± NA | 47.705 ± 21.696 | 8.333 ± 20.412 |
| Cosine | *27.482 ± 10.570* | 8.333 ± 20.412 | *81.277 ± 13.038* | 0.000 ± NA |
| Euclidean | 26.272 ± 11.123 | 16.667 ± 40.825 | 68.837 ± 21.759 | *13.889 ± 22.153* |
| Euclidean PCA | 13.309 ± 8.928 | 0.000 ± NA | 36.872 ± 26.105 | 8.333 ± 20.412 |

| Phenotypic similarity method | Distinctness (%) *** | Distinctness across lines (%) ** | Distinctness MOAs (%) *** | Distinctness across lines MOAs (%) ** |
|---|---|---|---|---|
| TripletLoss | 8.992 ± 9.139 | 70.667 ± 22.745 | 31.216 ± 21.008 | 71.200 ± 22.487 |
| Rank | **17.271 ± 9.810** | 85.333 ± 18.451 | 57.130 ± 17.645 | 87.067 ± 18.507 |
| Zhang | *16.607 ± 9.591* | 86.000 ± 18.062 | *57.219 ± 17.879* | 86.400 ± 18.328 |
| Cosine | 12.272 ± 8.283 | *87.333 ± 16.519* | 55.101 ± 17.611 | *88.267 ± 16.603* |
| Euclidean | 7.326 ± 6.176 | 81.333 ± 16.546 | 44.966 ± 17.251 | 82.000 ± 16.971 |
| Euclidean PCA | 10.209 ± 6.641 | **91.333 ± 13.433** | **64.255 ± 15.331** | **92.400 ± 13.149** |

| Phenotypic similarity method | Uniqueness (AUROC) *** | Uniqueness across lines (AUROC) *** | Uniqueness MOAs (AUROC) *** | Uniqueness across lines MOAs (AUROC) *** |
|---|---|---|---|---|
| TripletLoss | *0.948 ± 0.008 | *0.840 ± 0.013 | *0.882 ± 0.015 | *0.835 ± 0.015 |
| Rank | 0.896 ± 0.009 | 0.661 ± 0.018 | 0.760 ± 0.023 | 0.658 ± 0.017 |
| Zhang | 0.913 ± 0.008 | 0.678 ± 0.025 | 0.771 ± 0.019 | 0.674 ± 0.023 |
| Cosine | *0.917 ± 0.011* | 0.688 ± 0.019 | *0.799 ± 0.024* | 0.686 ± 0.018 |
| Euclidean | 0.893 ± 0.010 | *0.719 ± 0.016* | 0.771 ± 0.028 | *0.717 ± 0.016* |
| Euclidean PCA | 0.881 ± 0.012 | 0.668 ± 0.009 | 0.731 ± 0.020 | 0.668 ± 0.011 |

Insights into population separation which may go unnoticed when judging just the singular percent replicating value calculated using the 95$^{th}$ percentile of the non-replicate distribution, we therefore provide plots of compactness as a function of percent replicating versus a range of percentile cutoffs for all replicate criteria and compound groups in the supporting information accompanying this manuscript (see Figures S11-S14, and Figures S19-S22).

In summary for all filtered CMap compounds, the TripletLoss-derived phenotypic similarity method was the top performing metric in 9 out of 12 benchmark tasks (significantly outperforming the next top ranked metric in 7 out of 12), dominating the compactness and uniqueness task but being outperformed by the Euclidean metrics in three distinctness benchmark tasks . Similar performance was seen for the JUMP MOA compound subset, with the TripletLoss-derived metric being ranked top in 7 out of 12 setups.

## DISCUSSION

Results from the compactness tasks show the TripletLoss metric outperforming other phenotypic similarity methods in all but one setup; 'compactness MOAs'. A similarly strong performance is noted in the uniqueness benchmark, for which the TripletLoss model is ideally suited as during model training an angular loss term applied to triplets (anchor, positive and negative) serves to maximally separate non-matching treatments from matching, increasing uniqueness. Distinctness results with all filtered CMap compounds show EuclideanPCA as the predominant significant performance metric across all replicate groups, with the TripletLoss similarity method only significant for the first replicate group criteria. Whilst a simple metric, applying PCA before making measurements in this space ensures that the embedding is set up in such a way that the maximum explainable variance is captured along each axis, with active compound treatments providing the variance and subsequent easy measurement and high benchmark scores of EuclideanPCA. It is predictable that angular-only metrics like TripletLoss and Cosine would perform badly in this benchmark due to behavior when encountering noise within two closely positioned or overlapping groups where small variations can give rise to large angular changes. Logically, and as demonstrated by the results, EuclideanPCA appears to be the best similarity method evaluated for HTS hit calling. The uniqueness benchmark task shows the TripletLoss method performing significantly better than the second top ranked method for all replicate group criteria, split evaluations, and compound sets. For a simple model used to exemplify the split, training, and benchmarking capabilities of Leak Proof CMap, the approach performs exceptionally well.

Leak Proof CMap establishes a leak proof set of CMap splits for training and rigorous evaluation of L1000 phenotypic similarity methods, enabling fair comparison of new phenotypic similarity method performance. The trained TripletLoss metric achieved split averaged uniqueness (AUROC) task scores of 0.916 ± 0.011, followed closely by the Zhang metric with a score of 0.852 ± 0.020. Performance comparisons to literature techniques are not possible due to the use of different and often unavailable test sets.

## CONCLUSIONS

We have established, and used the Leak Proof CMap package in application of what we believe to be the most rigorous leak proof data splitting regime publicly reported on the CMap database or any L1000 data. We used these splits and inbuilt functionality to train a new simple triplet loss-derived phenotypic similarity method in a regime aiming to closely simulate training on known cell lines and MOAs, and testing on novel, unseen cell lines and MOAs. Benchmarking of this new method along with a selection of traditional and CMap/L1000 tailored phenotypic similarity methods using the carefully chosen compactness, distinctness and uniqueness benchmark tasks highlights the improvements possible with the application of AI/ML-derived phenotypic similarity methods using even relatively simple triplet-loss based approaches. Application of more up-to-date state of the art techniques would no doubt produce performant methods able to more thoroughly exploit information captured from L1000 readouts and enable use of public data for MOA determination even in novel cell lines. With these improved methods established, it may be envisioned that the L1000 technology takes a more prominent role in phenotypic discovery pipelines. Whilst L1000 is not a high throughput technique, it may be deployed after high throughput technologies such as high content imaging[70] (HCI) and used as a secondary confirmatory assay. Literature demonstrates that transcriptomic responses can be estimated from HCI[71], however L1000 may be particularly well suited for use alongside imaging as highlighted in a study by Way et.al. finding that complementary information was captured by the two techniques[57]. Additionally, L1000 is not impacted by many issues encountered in HCI such as difficulty in detecting morphological changes for ~50 % of human gene perturbations[72,73], and poor standardization caused by different acquisition instrumentation, dyes and cell treatments used[74]. There is hope however, that advances in AI/ML and optionally the integration[79-81] of additional data such as small molecule structure[75-77] will address these issues and greatly enhance HCI-based MOA assignment[78]. These methods mechanistically inform the selection of hit molecules in early stage drug discovery and can be positioned to support triage and downstream target deconvolution of hit from phenotypic screening assays as well as evaluation of the specificity or polypharmacology of hits identified from target based screening.

With continued improvement of metrics driven by standardized benchmarking, we hope to see increased use of the L1000 assay technology in roles alongside higher throughput phenotypic primary screening assays, opening up areas of valuable therapeutic novelty and efficacy, in the fight against rapidly dwindling drug approval rates relative to research and development expenditure.

## SUPPORTING INFORMATION

Leak_Proof_CMAP_supporting_information - Contains supporting mode hyperparameter table and model performance figures including boxplots and percent replicating plots over a range of percentile cut-offs (.DOCX)

Leak_Proof_CMAP_supporting_information – Contains output from hyperparameter optimization of the TripletLoss metric, P values for tests conducted on the percent replicating, percent distinct, and AUROC benchmarks, along with split information including composition by cell lines and MOAs, and MOA counts for each cell line (.XLSX)

## ACKNOWLEDGMENTS

This work has been supported by the UKRI, MRC Research Grant MR/W003996/1 "Accelerating medicine development timelines through new approaches in knowledge extraction from large biological datasets". Development of material in this manuscript has been supported by GSK

Conflict of Interest: none declared.

# Supporting information
# Leak Proof CMAP; a framework for evaluation of cell line agnostic L1000 similarity methods.

Steven Shave[1,2], Richard Kasprowicz[2], Abdullah M Athar[3], Denise Vlachou[2], Neil O Carragher[1], and Cuong Q Nguyen[4*]

[1]Edinburgh Cancer Research, Cancer Research UK Scotland Centre, Institute of Genetics and Cancer, University of Edinburgh, Crewe Road South, Edinburgh, EH4 2XR, UK. Email: neil.carragher@ed.ac.uk

[2]GSK Medicines Research Centre, Stevenage, SG1 2NY, UK.

[3]Artificial Intelligence and Machine Learning, GSK, The Stanley Building, London, N1C 4AG, UK.

[4]Artificial Intelligence and Machine Learning, GSK, 5th Floor, Suite 1, 259 E. Grand Ave, South San Francisco, California, 94080, USA. Email: cuong.q.nguyen@gsk.com

* Corresponding author, to whom correspondence should be addressed.

# Hyperparameter scanning and model selection.

*Table S3 – Hyperparameters from Optuna trial 155 achieving the highest encountered validation accuracy of 0.8728. Model used split 1,1 training data for training and validation data for calculation of validation accuracy.*

| Hyperparameter | Value |
|---|---|
| **Network shape** | [956, 637, 916, 128] |
| **Batch size** | 2466 |
| **Dropout rate** | 0.332153 |
| **Learning rate** | 0.027225 |

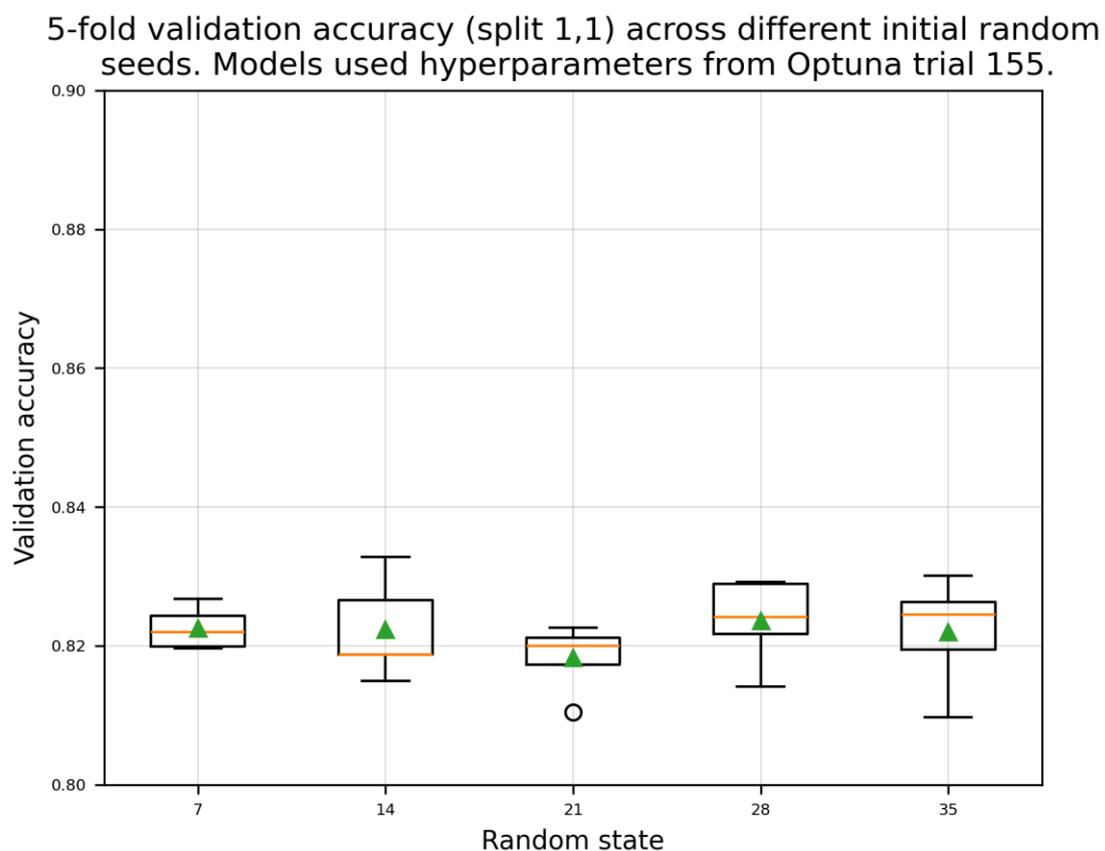

*Figure S1 – 5-fold validation accuracy across 5 different initial random states for models trained using the hyperparameters of Optuna trial 155 and split 1,1. All initial random states produce performant models in 5-fold cross validation with no significant performance differences identified using a one-way ANOVA test (P value = .69). Boxplots capture 5-fold validation accuracy scores for each initial starting seed (7, 14, 21, 28, 35). Orange bars denote median scores, and green triangles mean scores. Whiskers are 1.5 times the interquartile range (box body) below quartile 1 and above quartile 3. Outliers are anything beyond whiskers.*

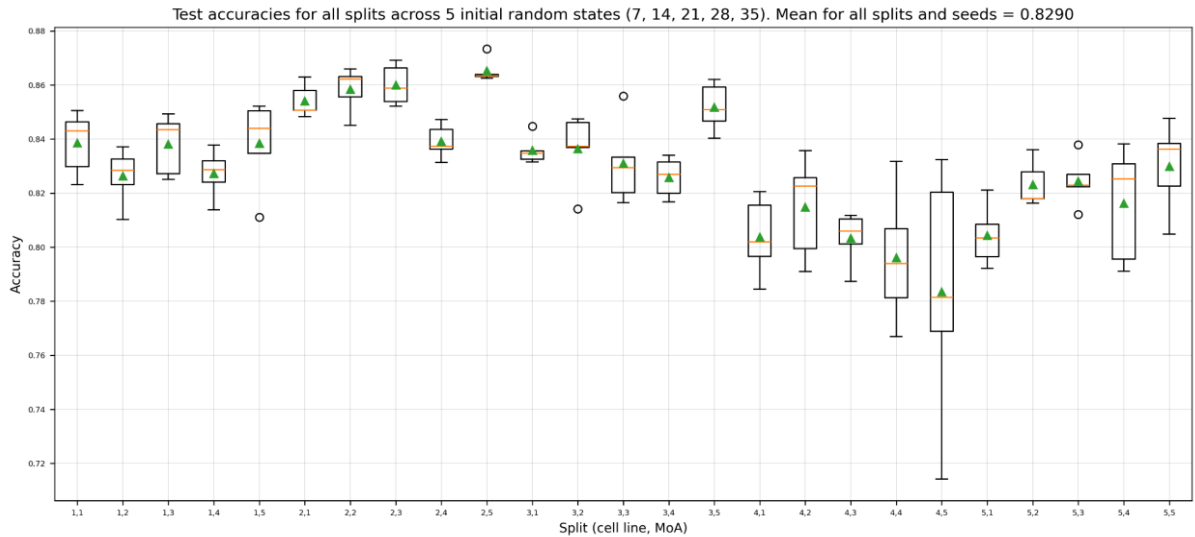

*Figure S2 – Test accuracies for all splits achieved with models trained on merged training and validation data (from their own split) before being evaluated using their associated held out test data. All models use hyperparameters from Optuna trial 155 and are trained using 5 different initial random states; 7, 14, 21, 28, and 35. Training with different initial random states and different splits demonstrates that model/trial 155 is stable and performant across all splits. Whilst test sets were evaluated with multiple seeds, a decision to use a random seed of 7 for subsequent phenotypic similarity method evaluation was taken before generating this data so as not to bias performance. Boxplots capture test accuracies for each split across different initial random states. Green triangles represent means and orange bars medians across initial starting states. Whiskers are 1.5 times the interquartile range (box body) below quartile 1 and above quartile 3. Outliers are anything beyond whiskers.*

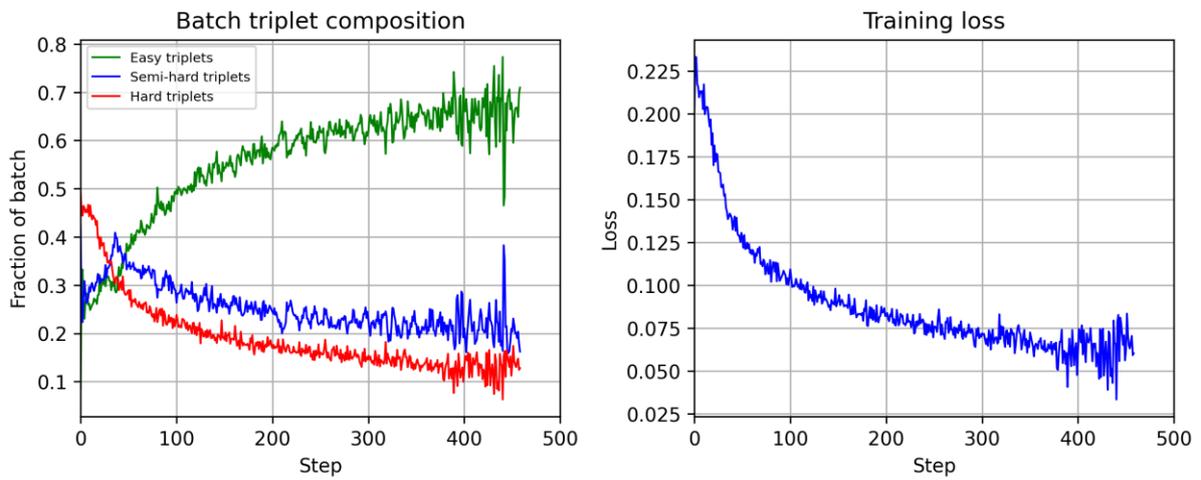

*Figure S3 – Fraction easy, semi-hard, and hard triplets encountered during training (left), shown alongside training loss (right). Triplets were monitored during training of a model using hyperparameters from Optuna trial 155 on split 1,1 training data. Early in model training (< 50 steps), we see hard triplets becoming either semi hard or easy triplets. As training progresses, the fraction of semi-hard triplets begins to reduce, with both hard and semi-hard triplets becoming easy triplets. This behavior appears to plateau for semi-hard triplets at around 200 steps and remain constant at a fraction around 0.22. Reduction of hard triplets appears to plateau later at a fraction around 0.15 after 300 steps. The triplet forming scheme ignoring cell line and concentration appears to naturally form semi-hard triplets which along with hard triplets, shift towards becoming easy triplets during training, after which the model may easily discriminate between triplet anchors, positives, and negatives.*

## Compactness vs Distinctness vs Uniqueness

The three tasks dubbed 'compactness' (evaluated using the percent replicating metric), 'distinctness' (evaluated using permutation testing), and 'uniqueness' (evaluated using the AUROC metric) can be used to benchmark and characterize phenotypic similarity methods, each looking at different aspects of their performance. The theoretically ideal phenotypic similarity method would achieve:

- high compactness scores - indicating that replicates are close together in phenotypic space.
- high distinctness scores - indicating that it can detect active verses inactive compounds (hit calling).
- high uniqueness scores - indicating that treatment replicates are highly unique in that no other non-replicate treatments occupy nearby or overlapping phenotypic space.

It is intuitive that compactness impacts uniqueness as large replicate spreads occupy more phenotypic space and therefore are more likely to overlap with non-replicates, decreasing AUROC scores. Similarly, perfectly compact replicate treatments occupying the same phenotypic space would most highly rank their replicates as most similar and achieve perfect AUROC scores of 1 (unless sharing the exact same space with non-replicate treatments). The same cannot be said for uniqueness impacting compactness as overlapping replicate and non-replicate treatments resulting in low AUROC scores does not impact replicate compactness.

The Jupyter notebook named 'leak_proof_cmap_07_uniqueness_vs_compactness_treatments.ipynb' within the Leak Proof CMAP source repository explores the relationship between scores obtained in these three tasks.

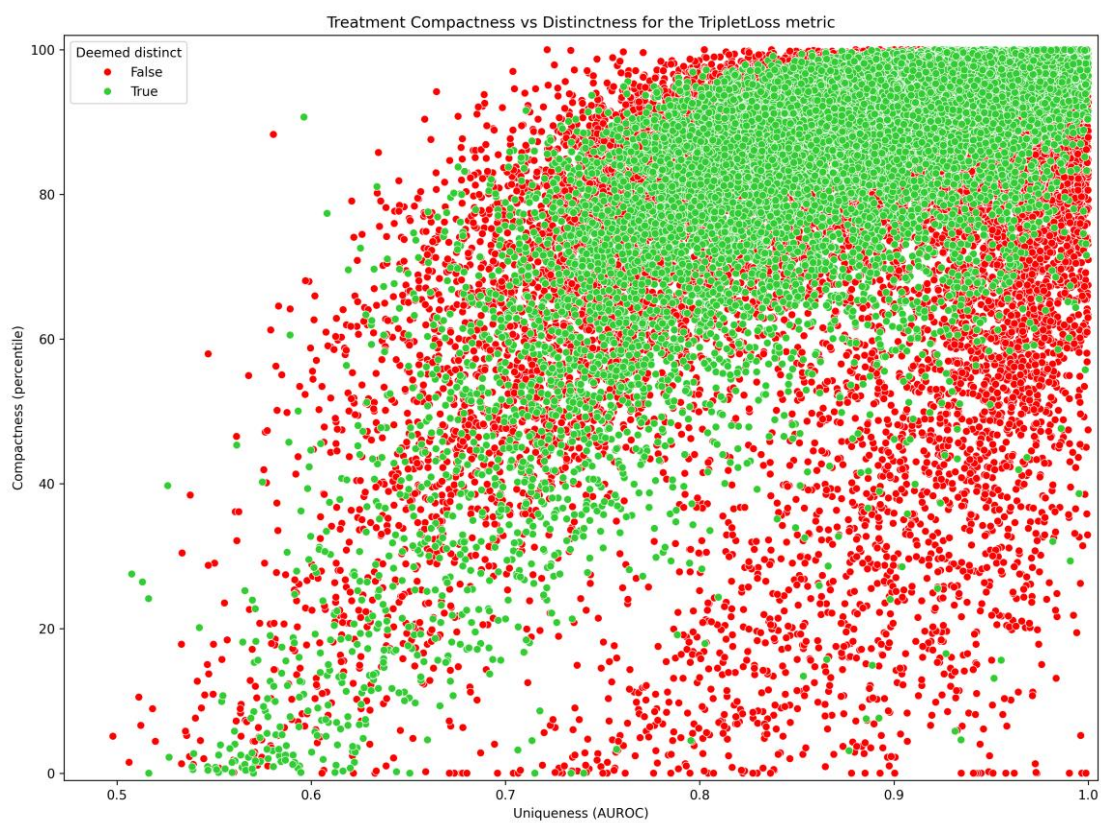

*Figure S4 Compactness vs Uniqueness for the TripletLoss metric, colored by Distinctness (green for* P *values < 0.05, red otherwise) for replicate treatments matching treatment, dose, and cell line. A high correlation between Compactness and Uniqueness is evident (r= 0.618), with a negligible correlation between Distinctness and Compactness (0.212) and Distinctness and Uniqueness (r= 0.166).*

*Figure S4 shows compactness versus uniqueness for the TripletLoss metric applied to Leak Proof CMAP filtered compounds with the color capturing the Distinctness (from DMSO) of the replicate group, green for distinct and red for non-distinct, assigned via permutation testing at the 0.05 level. Replicates match treatment, dose, and cell line. Figure S4 appears to show a high correlation between Compactness and Uniqueness. Pearson correlation coefficients are shown in table S2.*

*Table S4 – Pearson correlation coefficient matrix for Distinctness, Uniqueness, and Compactness of replicate treatments as achieved using the TripletLoss metric, matching treatment, dose, and cell line, using the Cosine metric and all Leak Proof CMAP filtered compounds.*

|             | Compactness | Distinctness | Uniqueness |
|---|---|---|---|
| **Compactness**  | 1.000 | 0.212 | 0.618 |
| **Distinctness** | 0.212 | 1.000 | 0.166 |
| **Uniqueness**   | 0.618 | 0.166 | 1.000 |

Slightly different correlations are observed using other metrics such as EuclideanPCA shown below in *Figure S5*, and the Pearson correlation coefficient matrix shown in Table S3, with similar correlations between compactness and uniqueness, but with distinctness moderately negatively correlated to compactness, and weakly negative to uniqueness.

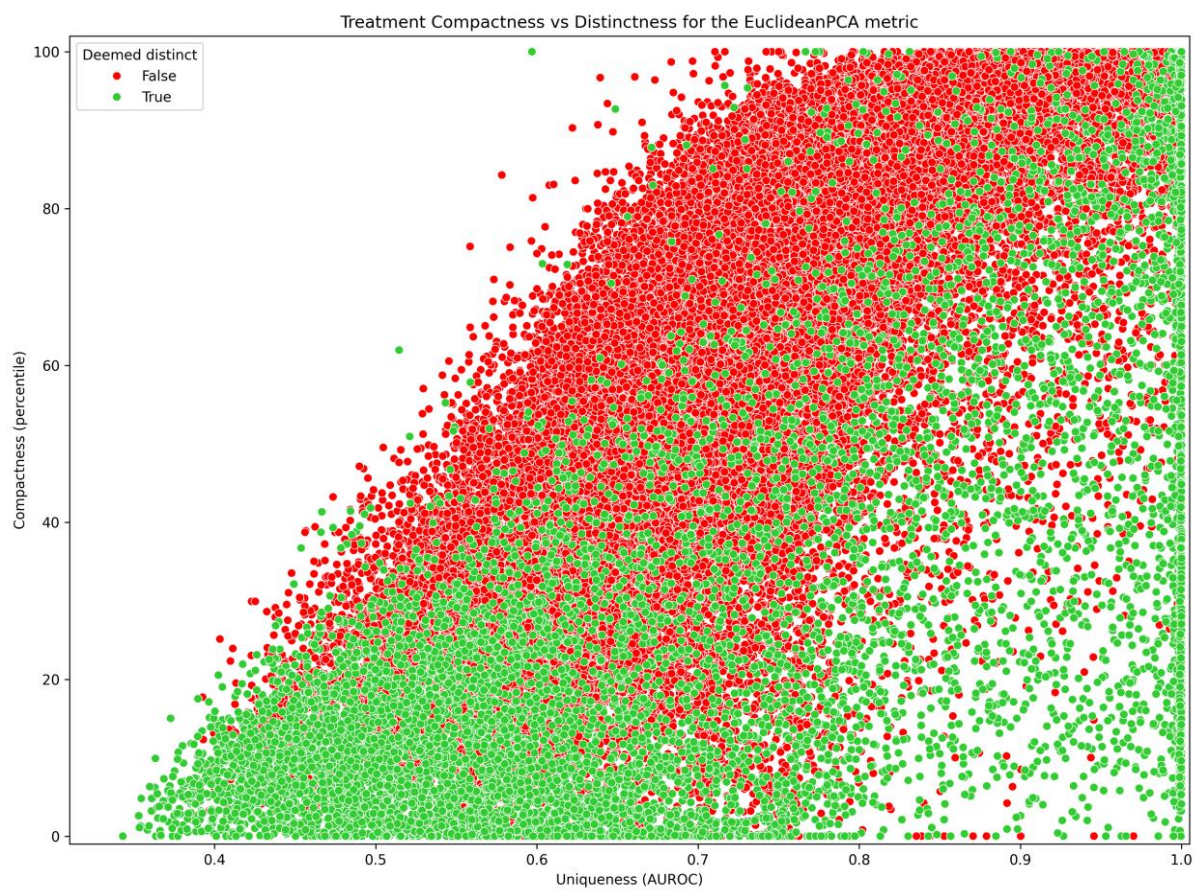

*Figure S5 - Compactness vs Uniqueness for the EuclideanPCA metric, colored by Distinctness (green for P values < 0.05, red otherwise) for replicate treatments matching treatment, dose, and cell line. A high correlation between Compactness and Uniqueness is evident (r= 0.618), with a negligible correlation between Distinctness and Compactness (0.212) and Distinctness and Uniqueness (r= 0.166).*

*Table S5 – Pearson correlation coefficient matrix for Distinctness, Uniqueness, and Compactness of replicate treatments as achieved using the EuclideanPCA metric, matching treatment, dose, and cell line, using the Cosine metric and all Leak Proof CMAP filtered compounds.*

|              | Compactness | Distinctness | Uniqueness |
|--------------|-------------|--------------|------------|
| **Compactness**  | 1.000       | -0.537       | 0.708      |
| **Distinctness** | -0.537      | 1.000        | -0.132     |
| **Uniqueness**   | 0.708       | -0.132       | 1.000      |

Distinctness depends on only the relative positioning of replicate groups and DMSO, it is therefore logical that it is weakly correlated with Compactness and Uniqueness (as demonstrated for the TripletLoss metric). However, we can imagine simulating a replicate group and increasing the spread until it occupies vast portions of phenotypic space including the space of DMSO, at which point distinctness is lost. Whilst this is an extreme example, it illustrates that a borderline distinctness assignment for a replicate group with a P value of around 0.05 may be heavily influenced by changes in group spread.

We may think about treatments occupying the four extreme corners of the plot in *Figure S5* as having the following characteristics visualized in *Figure S6*:

- High Compactness and High Uniqueness (Top right of *Figure S5*)
    - The replicate group is highly compact and occupies an area of phenotypic space in which the applied metric measures the closest treatment to all replicate group treatments to be treatments from the replicate group.
- Low Compactness and High Uniqueness (Bottom right of *Figure S5*)
    - The replicate group is spread extremely widely, but occupies an area of phenotypic space so distant from other treatments that the applied metric measures the closest treatment to all replicate group treatments to be treatments from the replicate group. Even with the extremely wide spread of the replicate group, not enough space is spanned to encroach on the phenotypic space of other treatments.
- Low Compactness and Low Uniqueness (Bottom left of *Figure S5*)
    - A poorly performing replicate group which is widely spread and spans an area of phenotypic space occupied by non-replicate treatments which are closer to replicates than all other replicates.
- High Compactness and Low Uniqueness (Top left of *Figure S5*)
    - The replicate group is highly compact, but this small area of phenotypic space is also occupied by other non-replicate treatments.

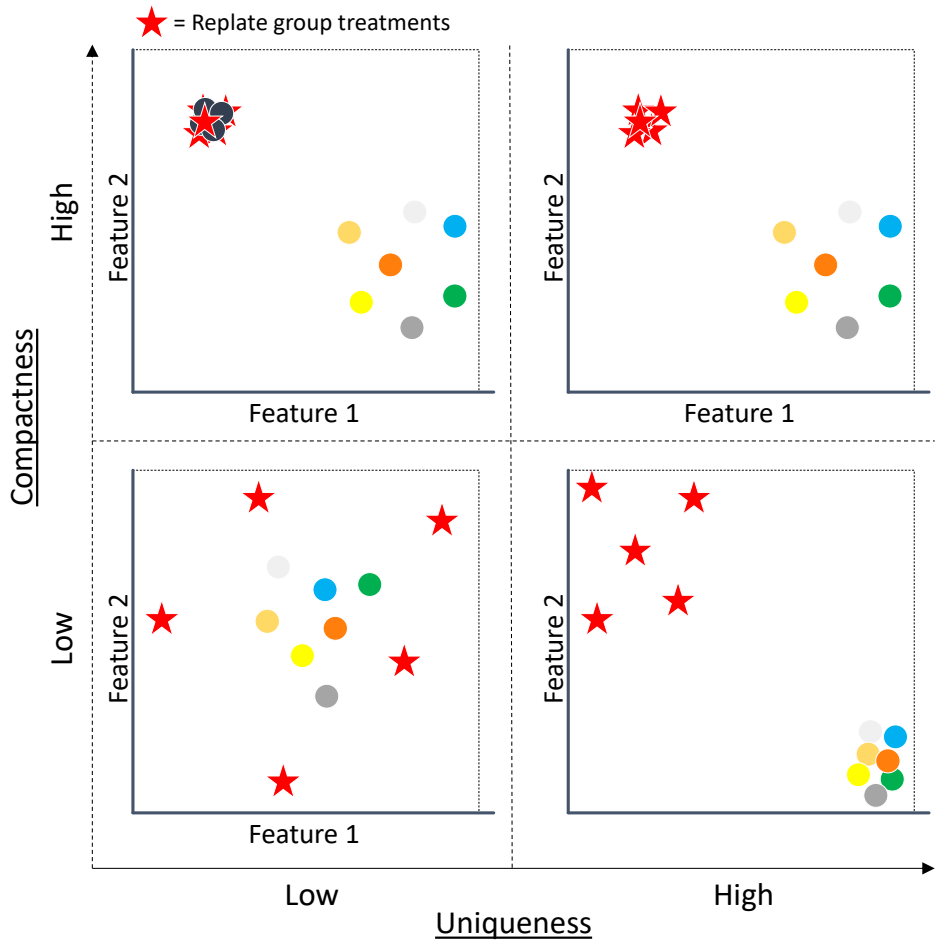

*Figure S6 – Mockup visualization of high/low Compactness vs high/low Uniqueness for a replicate group (red star) amongst a background of other treatments (colored circles).*

Extracting treatments at these extreme corners allows us to visualize the compactness, distinctness and uniqueness landscape.

## High Compactness and High Uniqueness (Top right of Figure S5)

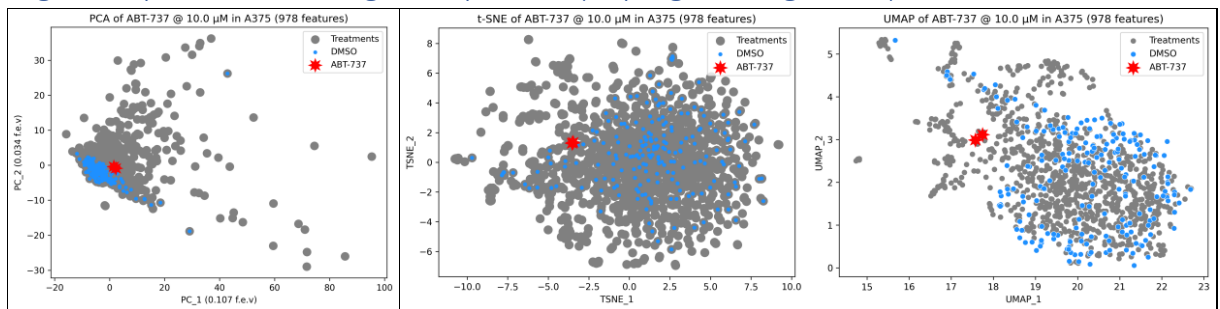

*Figure S 7 – High compactness and high uniqueness of treatment ABT-737 (red stars) at 10 μM in the A375 cell line against a background of different treatments (grey dots), and DMSO (blue dots). Although the 2D PCA (Left) explains only around 14 % of dataset variance, the replicates are compact and outside of the DMSO cloud, hinting at the highly compact nature treatment replicate groupings. t-SNE and UMAP plots (centre and right respectively) show highly compact treatments with their nearest neighbours being other replicates (high uniqueness).*

## Low Compactness and High Uniqueness (Bottom right of Figure S5)

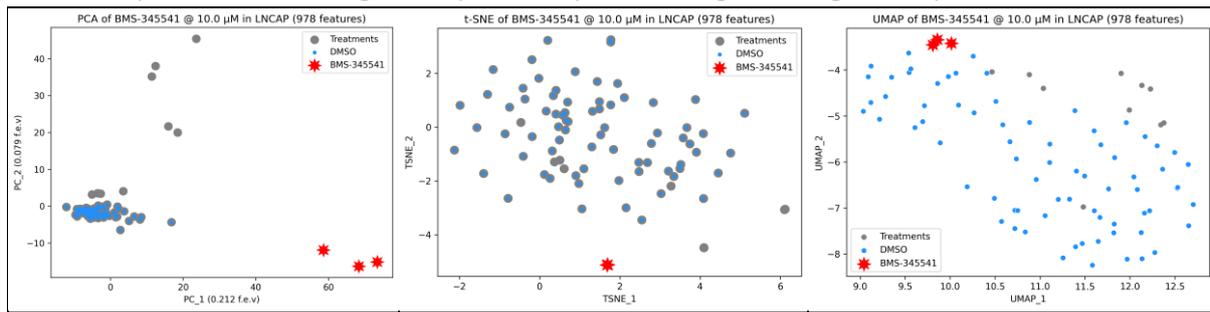

*Figure S 8 - Low compactness and high uniqueness of treatment BMS-345541 (red stars) at 10 µM in the LNCAP cell line against a background of different treatments (grey dots), and DMSO (blue dots). Although the 2D PCA (Left) explains only around 29 % of dataset variance, the replicates are widely spread in comparison to background (low compactness) are far away from the DMSO cloud and all other replicate treatments indicating high distinctness and high uniqueness respectively. t-SNE and UMAP plots (centre and right respectively) show replicate treatments overlapping or very close to each other indicating that their nearest neighbours are themselves regardless of spread, leading to high uniqueness.*

## Low Compactness and Low Uniqueness (Bottom left of Figure S5)

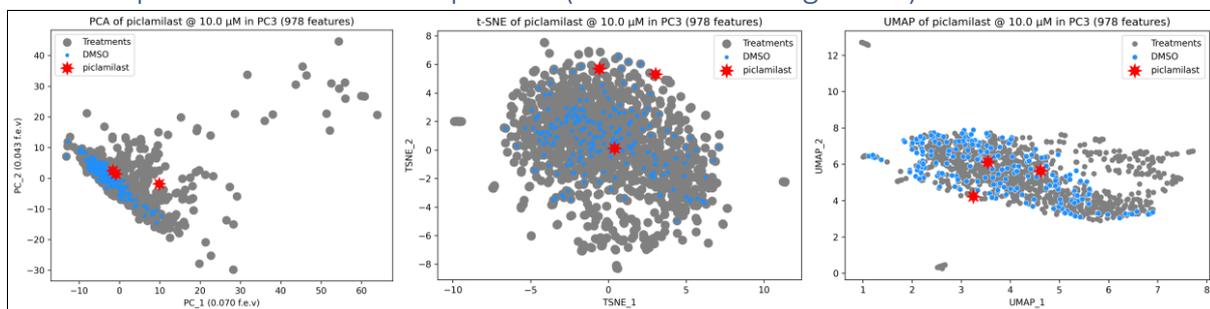

*Figure S 9 - Low compactness and low uniqueness of treatment piclamilast (red stars) at 10 µM in the PC3 cell line against a background of different treatments (grey dots), and DMSO (blue dots). Although the 2D PCA (Left) explains only around 11 % of dataset variance, the replicates have a high spread in comparison to background (low compactness). t-SNE and UMAP plots (centre and right respectively) show replicates spread, with different neighbors indicating low uniqueness.*

## High Compactness and Low Uniqueness (Top left of Figure S5)

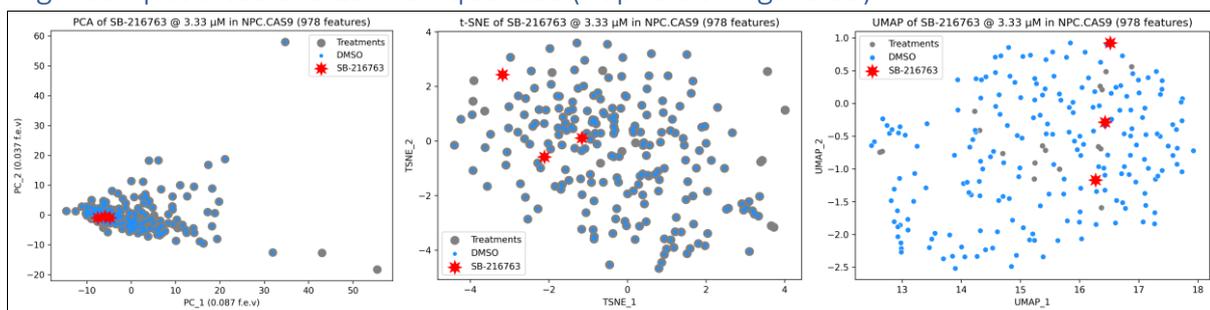

*Figure S 10 - High compactness and low uniqueness of treatment SB-216763 (red stars) at 3.33 µM in the NPC.CAS9 cell line against a background of different treatments (grey dots), and DMSO (blue dots). Whilst no truly extreme high compactness and low uniqueness treatments are evident in the scatterplot of compactness vs uniqueness (Figure S5), the most extreme "top left" treatment was used, equating to a compactness of 61.962 and a uniqueness of 0.515. Although the 2D PCA (Left) explains only around 12 % of dataset variance, the replicates appear compact in comparison to background. t-SNE and UMAP plots (centre and right respectively) show replicates spread, with different neighbors indicating low uniqueness.*

# Merged splits, results table

*Table S6 – Merged split benchmark results for compactness (percent replicating), distinctness from DMSO/hit calling (percent distinct), and retrieval of matching profiles (AUROC). For percent replicating and percent distinct, assignment of replicating or distinct was performed for all treatment groups within each split. These results across the 25 splits were then used to calculate a percent replicating/distinct. The same procedure was carried out for AUROC, with average AUROC scores for all treatment groups used to form a concatenated list comprising results from all splits, from which an average AUROC was calculated, along with standard deviations. Bold indicates phenotypic similarity methods with the highest score within each benchmark setup. A \* denotes that the top ranked similarity method significantly outperforms the second ranked with a P value of < .05, \*\* denotes a P value of < .01, and \*\*\* denotes a P value < .001*

### All filtered CMAP compounds

| Phenotypic similarity method | Compactness (%) | Compactness across lines (%) | Compactness MOAs (%) | Compactness across lines MOAs (%) |
|---|---|---|---|---|
| TripletLoss | \*\*\*54.809 | \*\*\*24.663 | 67.587 | \*\*\*19.619 |
| Rank | 37.227 | 12.742 | 77.521 | 9.905 |
| Zhang | 36.620 | 11.199 | 76.794 | 9.238 |
| Cosine | 35.676 | 10.411 | \*\*\*85.273 | 8.762 |
| Euclidean | 12.604 | 8.637 | 28.073 | 7.714 |
| Euclidean PCA | 11.383 | 5.846 | 27.873 | 4.762 |

| Phenotypic similarity method | Distinctness (%) | Distinctness across lines (%) | Distinctness MOAs (%) | Distinctness across lines MOAs (%) |
|---|---|---|---|---|
| TripletLoss | \*\*\*45.298 | 94.006 | 75.788 | 92.870 |
| Rank | 16.810 | 91.131 | 81.041 | 92.917 |
| Zhang | 16.157 | 90.443 | 80.465 | 92.407 |
| Cosine | 13.885 | 97.599 | 83.242 | \*\*\*97.222 |
| Euclidean | 19.760 | 95.260 | 81.766 | 94.074 |
| Euclidean PCA | 14.649 | \*\*\*97.951 | \*\*\*88.844 | 97.037 |

| Phenotypic similarity method | Uniqueness (AUROC) | Uniqueness across lines (AUROC) | Uniqueness MOAs (AUROC) | Uniqueness across lines MOAs (AUROC) |
|---|---|---|---|---|
| TripletLoss | **0.914 ± 0.084** | \*\*\*0.839 ± 0.051 | \*\*\* 0.773 ± 0.145 | \*\*\*0.732 ± 0.133 |
| Rank | 0.822 ± 0.141 | 0.681 ± 0.079 | 0.674 ± 0.114 | 0.622 ± 0.083 |
| Zhang | 0.849 ± 0.143 | 0.697 ± 0.082 | 0.678 ± 0.125 | 0.633 ± 0.088 |
| Cosine | 0.835 ± 0.143 | 0.686 ± 0.079 | 0.690 ± 0.121 | 0.627 ± 0.083 |
| Euclidean | 0.755 ± 0.155 | 0.579 ± 0.113 | 0.608 ± 0.148 | 0.551 ± 0.099 |
| Euclidean PCA | 0.761 ± 0.150 | 0.580 ± 0.089 | 0.594 ± 0.135 | 0.553 ± 0.078 |

### JUMP MOA compound subset

| Phenotypic similarity method | Compactness (%) | Compactness across lines (%) | Compactness MOAs (%) | Compactness across lines MOAs (%) |
|---|---|---|---|---|
| TripletLoss | \*\*\*35.876 | 0.000 | \*\*\*96.000 | 27.273 |
| Rank | 16.102 | 0.000 | 44.286 | 0.000 |
| Zhang | 15.960 | 0.000 | 44.000 | 9.091 |
| Cosine | 26.554 | 9.091 | 78.286 | 0.000 |
| Euclidean | 25.424 | 9.091 | 65.714 | 18.182 |
| Euclidean PCA | 12.994 | 0.000 | 34.857 | 9.091 |

| Phenotypic similarity method | Distinctness (%) | Distinctness across lines (%) | Distinctness MOAs (%) | Distinctness across lines MOAs (%) |
|---|---|---|---|---|
| TripletLoss | \*\*\*44.379 | 92.936 | 72.188 | 89.935 |
| Rank | 17.036 | 90.387 | 78.253 | 90.710 |
| Zhang | 16.386 | 89.717 | 77.651 | 90.151 |
| Cosine | 14.091 | 96.706 | 80.275 | 94.753 |
| Euclidean | 19.618 | 94.307 | 77.888 | 91.484 |
| Euclidean PCA | 14.789 | \*\*97.124 | \*\*85.868 | \*\*94.796 |

| Phenotypic similarity method | Uniqueness (AUROC) | Uniqueness across lines (AUROC) | Uniqueness MOAs (AUROC) | Uniqueness across lines MOAs (AUROC) |
|---|---|---|---|---|
| TripletLoss | \*\*\*0.949 ± 0.078 | \*\*\*0.840 ± 0.065 | \*\*\* 0.881 ± 0.086 | \*\*\*0.835 ± 0.065 |
| Rank | 0.896 ± 0.136 | 0.661 ± 0.092 | 0.758 ± 0.123 | 0.658 ± 0.095 |
| Zhang | 0.913 ± 0.137 | 0.678 ± 0.096 | 0.768 ± 0.139 | 0.674 ± 0.099 |
| Cosine | 0.918 ± 0.127 | 0.688 ± 0.100 | 0.798 ± 0.122 | 0.686 ± 0.103 |
| Euclidean | 0.893 ± 0.118 | 0.719 ± 0.090 | 0.770 ± 0.130 | 0.717 ± 0.090 |
| Euclidean PCA | 0.881 ± 0.129 | 0.668 ± 0.076 | 0.731 ± 0.140 | 0.668 ± 0.071 |

## Analysis of merged split results

Similar significances are seen for merged split and split averaged (manuscript Table 1) percent replicating performance using both compound sets with the notable exception of the significant outperformance of Cosine over Rank in percent replicating MOAs using all CMap filtered compounds (P value < .001). Good agreement is seen between merged split and split averaged performance in the percent distinct and AUROC benchmarks.

## Methods - McNemar's test

Binary classifications applied to treatments such as those intermediately calculated in the merged split percent replicating and distinct from DMSO benchmarks were evaluated using McNemar's test as implemented in the statsmodels (v 0.14.0) Python package, utilizing the binomial distribution for the test statistic. McNemar's test was applied to paired treatment outcome results comprising data from all test splits for two different phenotypic similarity methods. As only the top scoring method was compared against the second-best performing method, no multiple testing false discovery rate correction was required, although a full non-corrected and corrected (using the Benjamini-Hochberg multiple testing correction as implemented in the SciPy v 1.11.1 Python package) set of all-to-all P values are available in supporting information spreadsheet worksheets. Code used to apply this test is available in the LeakProofCMap source repository within Jupyter notebooks named 'leak_proof_cmap_03_compactness_eval.ipynb, and 'leak_proof_cmap_04_distinctness_eval.ipynb', evaluating percent replicating and distinct from DMSO benchmark performance respectively.

# Percent replicating plots across cutoffs
## All compounds
Split averaged plots

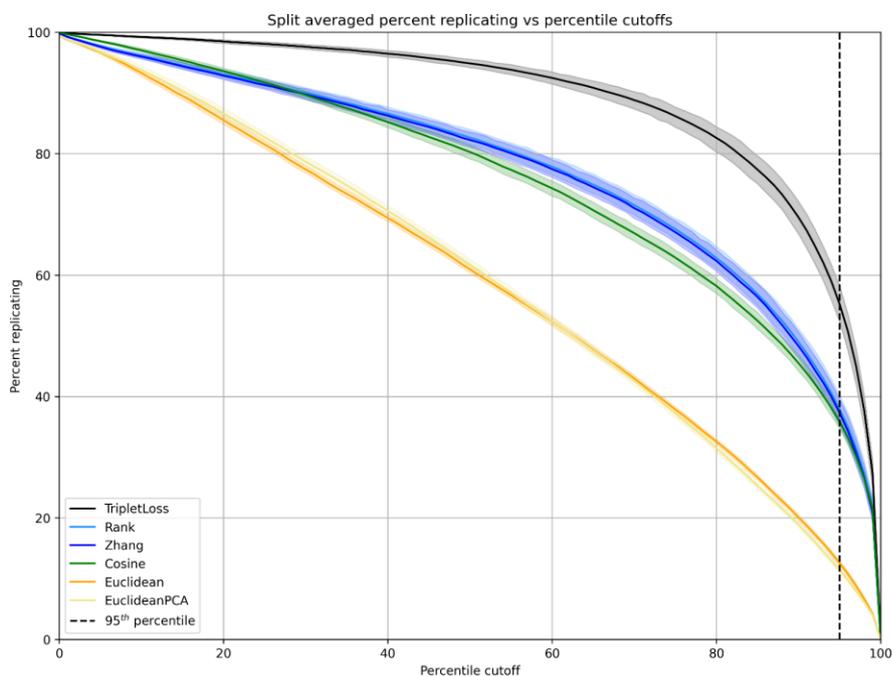

*Figure S11 –Percent replicating verses percentile cut-off for 25 test splits. Bands denote standard deviations of split scores.*

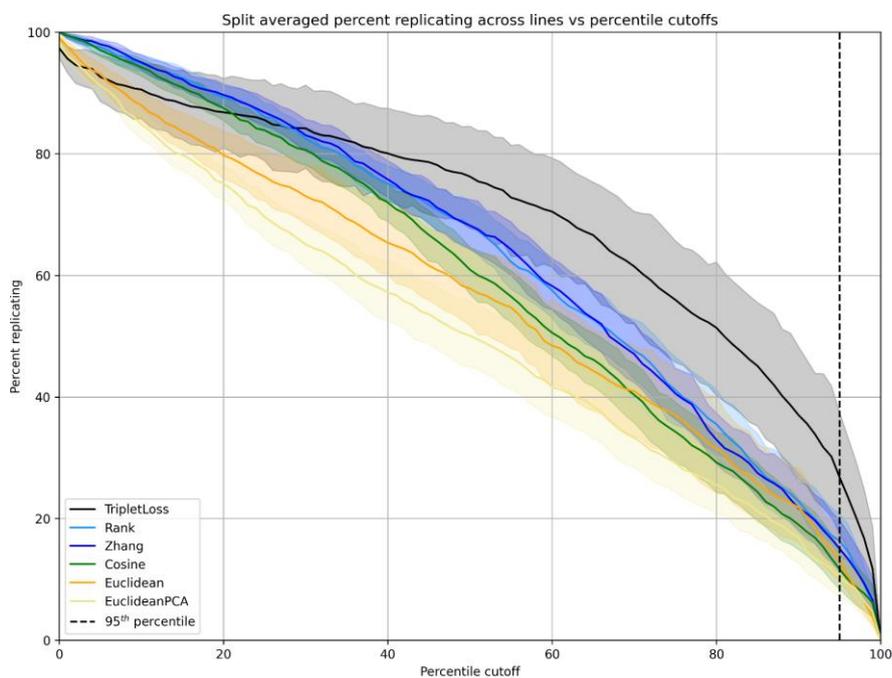

*Figure S12 –Percent replicating across lines verses percentile cut-off for 25 test splits. Bands denote standard deviations of split scores.*

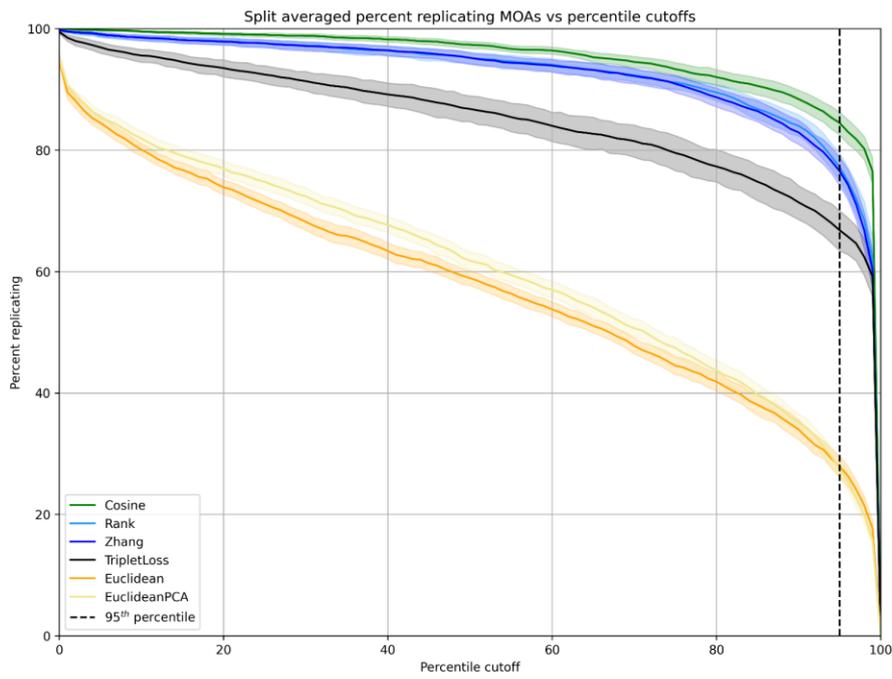

*Figure S13 –Percent replicating MOAs verses percentile cut-off for 25 test splits. Bands denote standard deviations of split scores.*

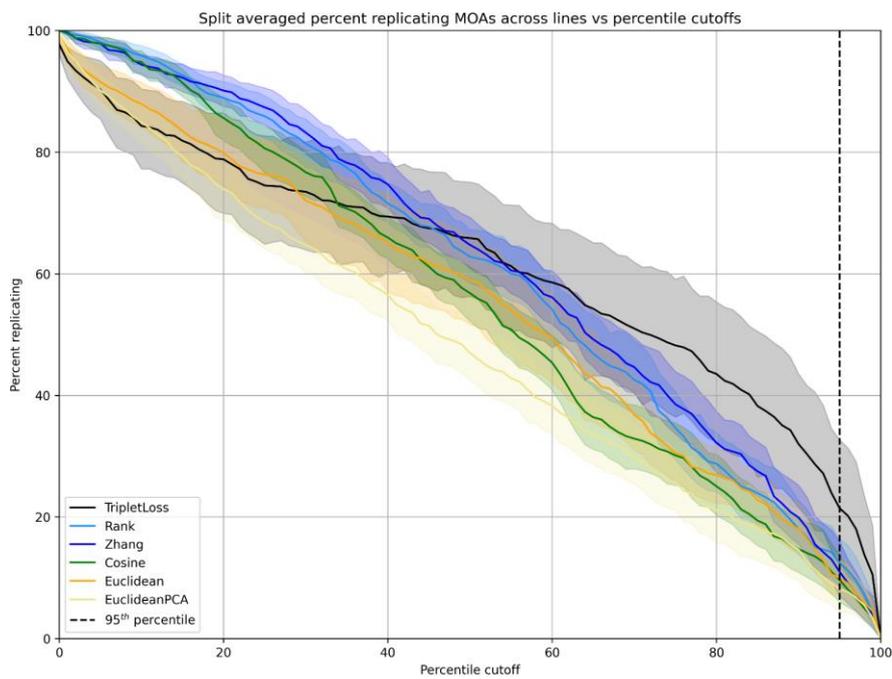

*Figure S14 –Percent replicating across lines MOAs verses percentile cut-off for 25 test splits. Bands denote standard deviations of split scores.*

# Boxplots at 95th percentile

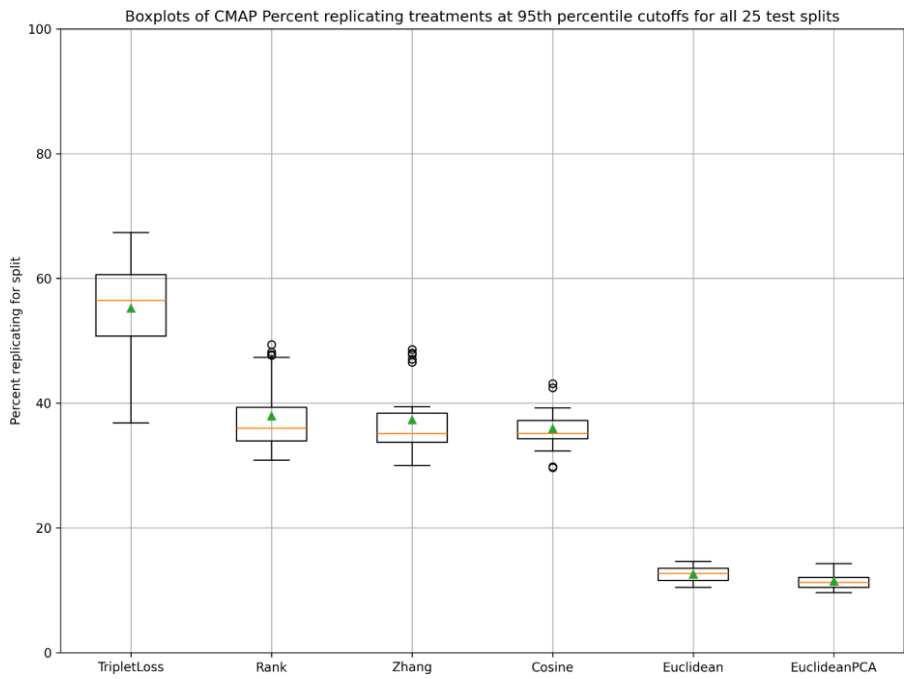

*Figure S15 – Split averaged percent replicating performance using the 95th percentile of the non-replicate distribution as the replication cut-off.*

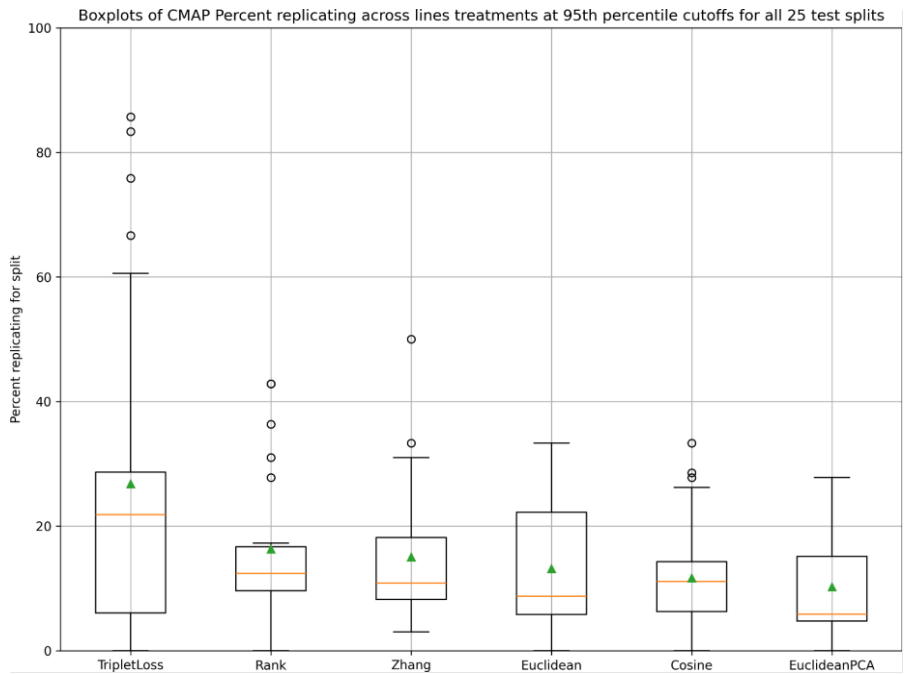

*Figure S16 – Split averaged percent replicating across lines performance using the 95th percentile of the non-replicate distribution as the replication cut-off.*

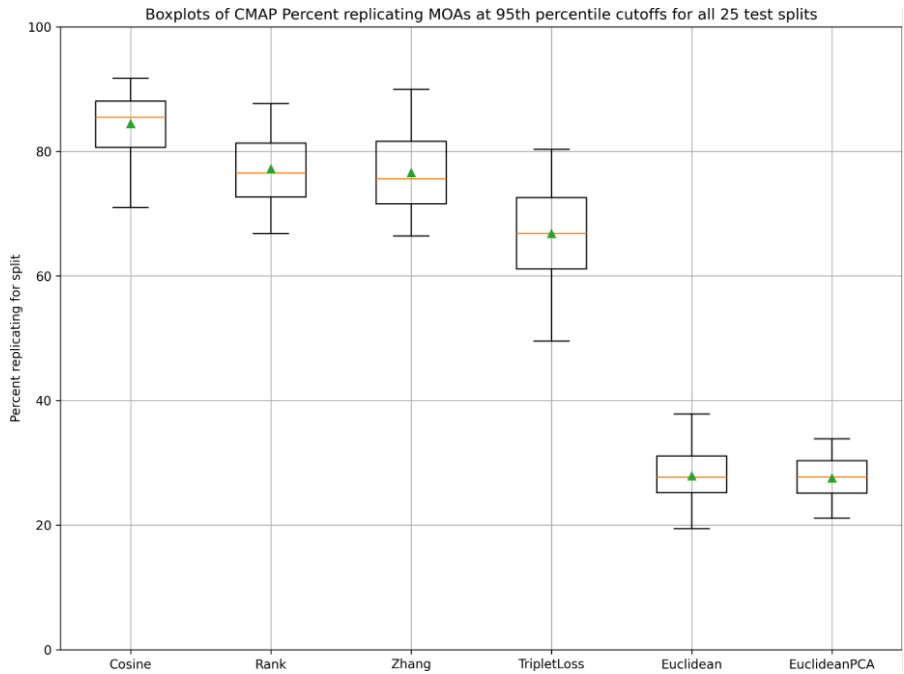

*Figure S17 – Split averaged percent replicating MOAs performance using the 95$^{th}$ percentile of the non-replicate distribution as the replication cut-off.*

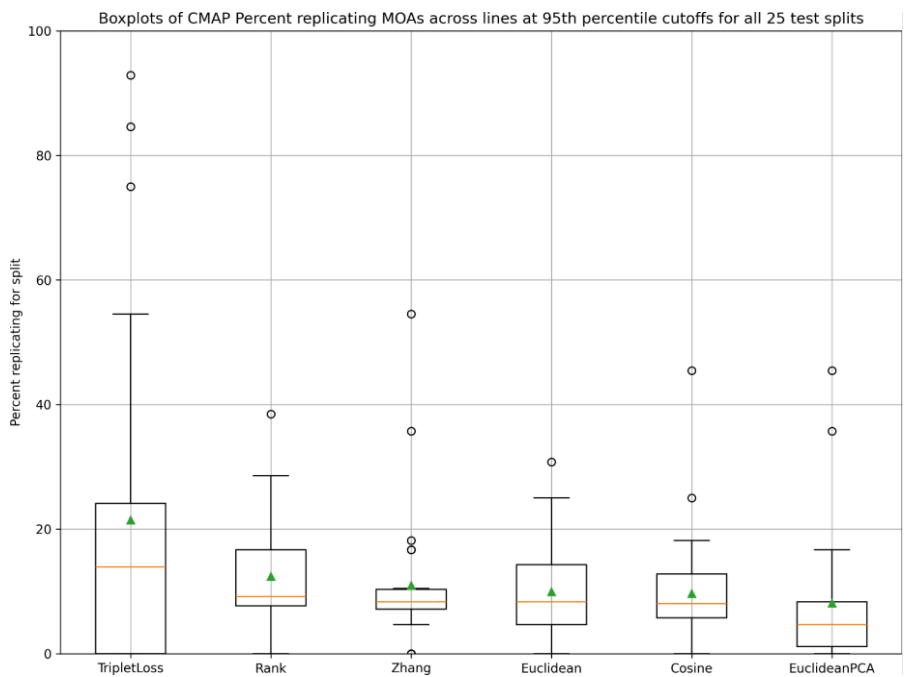

*Figure S18 – Split averaged percent replicating across lines MOAs performance using the 95$^{th}$ percentile of the non-replicate distribution as the replication cut-off.*

# JUMP MOA compound subset
## Split averaged plots

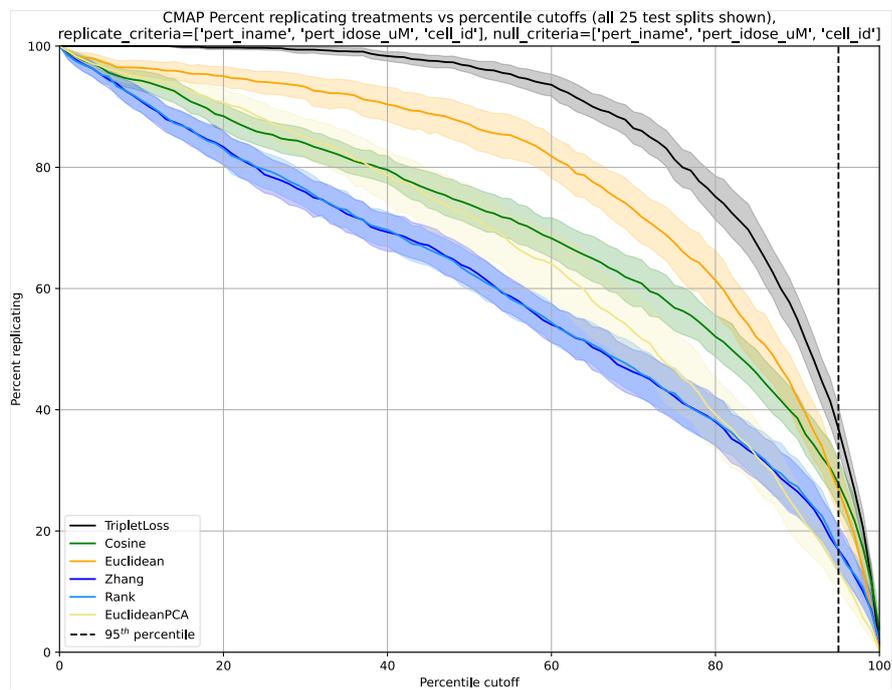

*Figure S19 – Split averaged percent replicating verses percentile cut-off. Bands denote standard deviations of split scores.*

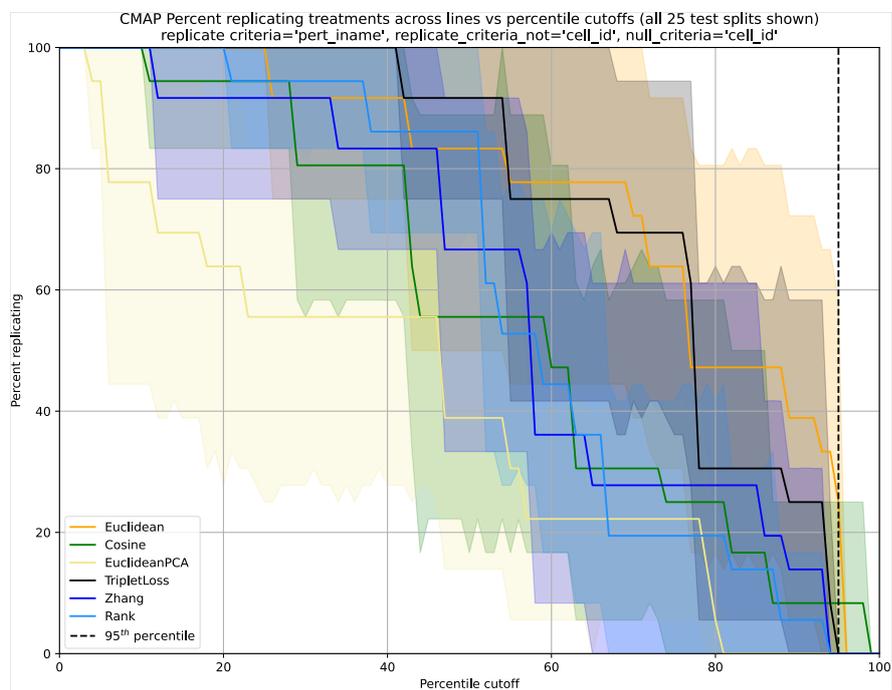

*Figure S20 – Split averaged percent replicating across lines verses percentile cut-off. Bands denote standard deviations of split scores.*

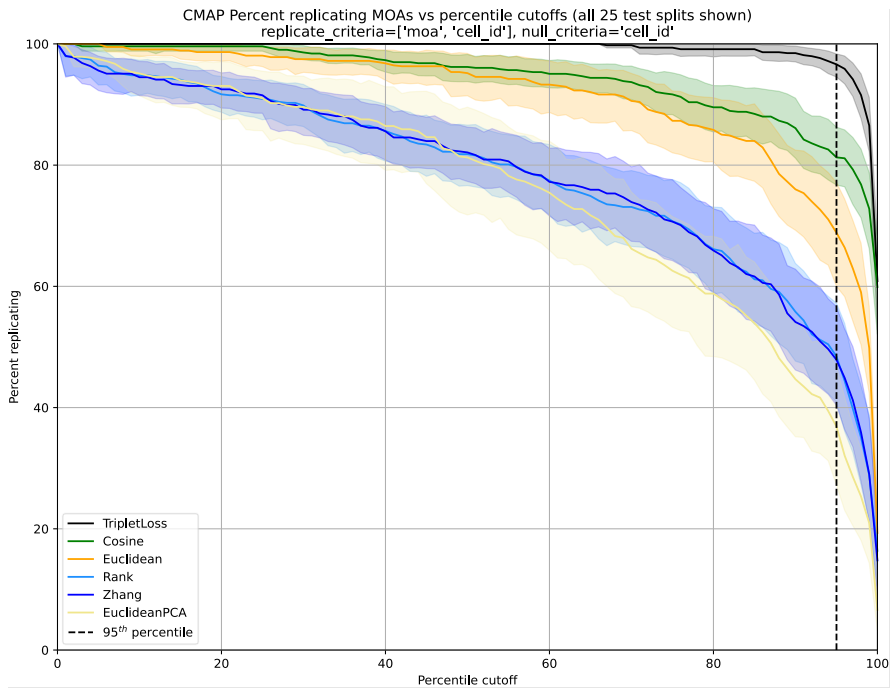

*Figure S21 – Split averaged percent replicating MOAs verses percentile cut-off. Bands denote standard deviations of split scores.*

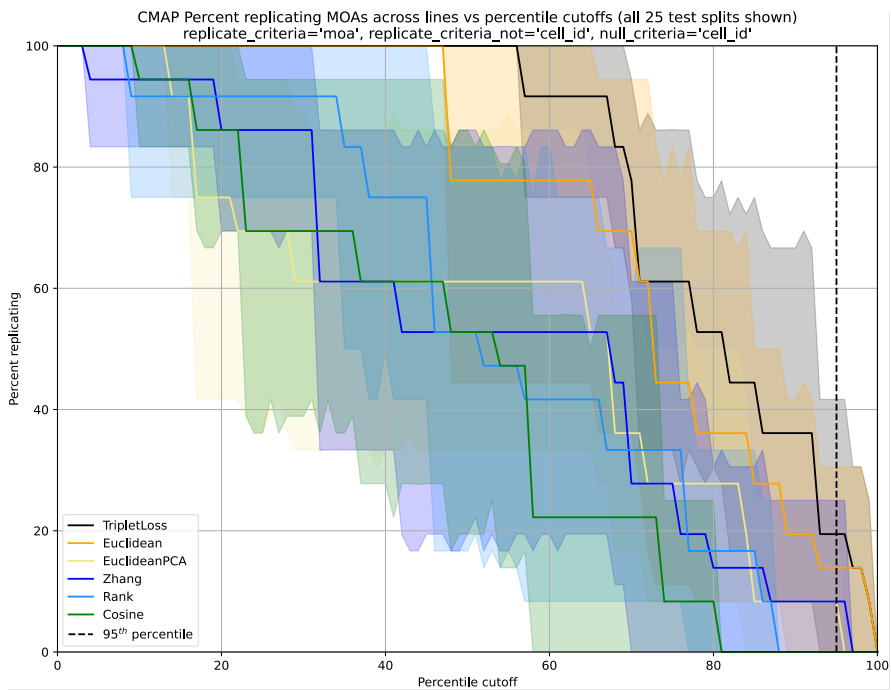

*Figure S22 – Split averaged percent replicating across lines MOAs verses percentile cut-off. Bands denote standard deviations of split scores.*

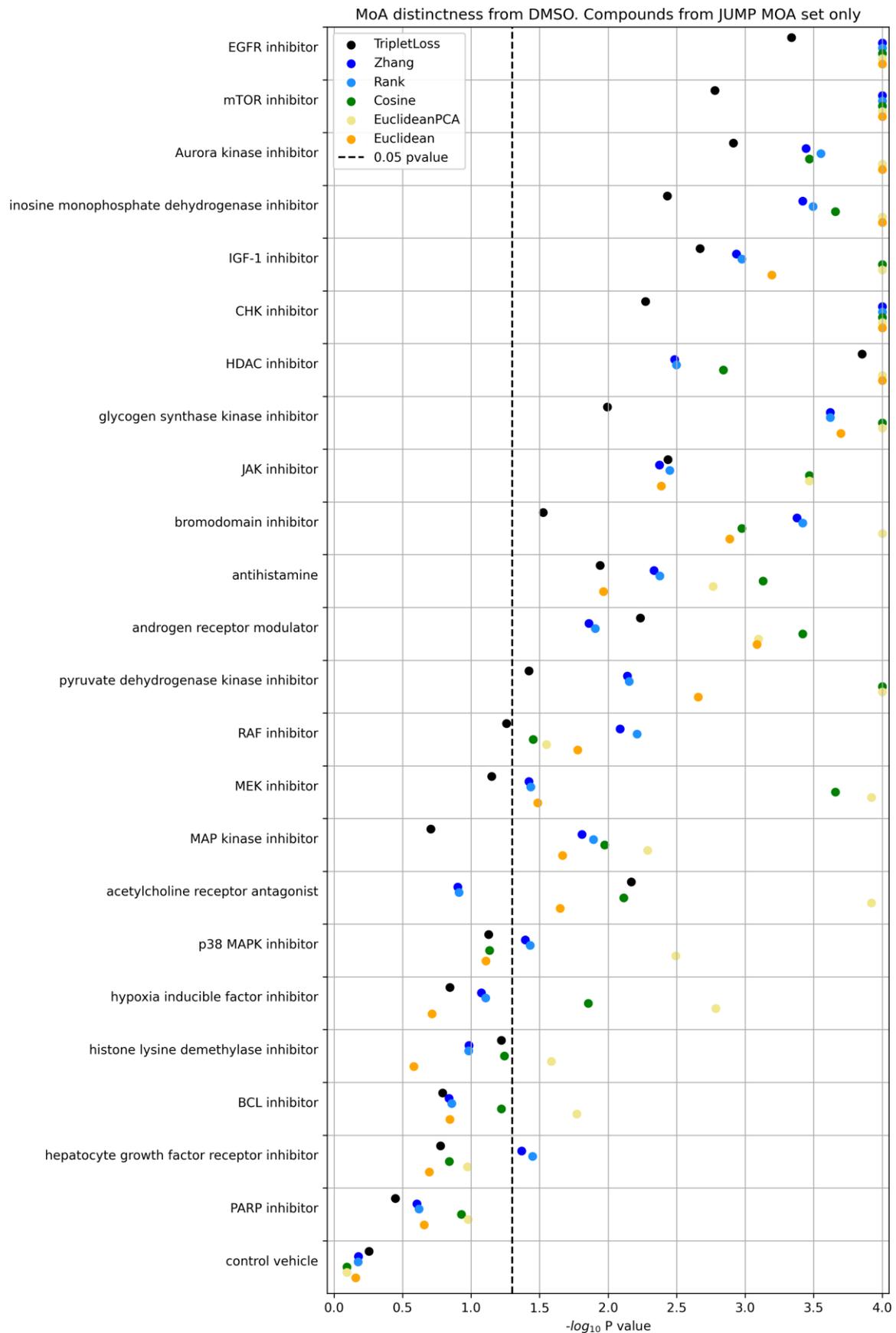

Figure S23 – MOA distinctness of JUMP MOA compounds from DMSO. Permutation test derived split averaged P values for treatments belonging to DMSO response set. Null hypothesis is that treatment and DMSO come from the same distribution.

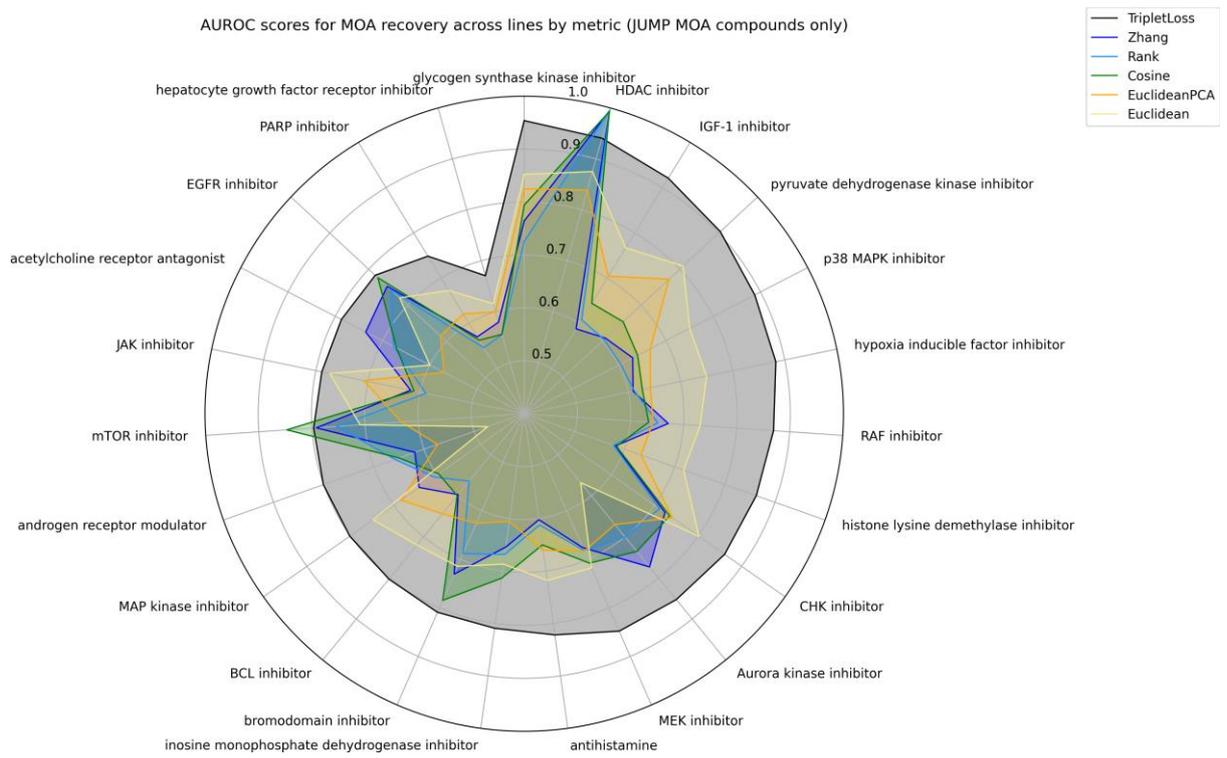

*Figure S24 – Radial plot of split averaged AUROC scores obtained in recall of JUMP MOA compounds across cell lines*